\documentclass[twocolumn,twocolappendix]{aastex63}%Apj 2columns

\pdfoutput=1 %for arXiv submissionok
\usepackage{amsmath,amstext}
\usepackage[T1]{fontenc}
\usepackage{apjfonts} 
\usepackage{comment}
\usepackage[figure,figure*]{hypcap}
\usepackage{multirow}
\usepackage{amsmath}
\usepackage{amsbsy}
\usepackage{nccmath}
\usepackage{rotating}

\usepackage{soul}

\setstcolor{red}
\newcommand{\rahms}[4]{$#1^{\rm h}#2^{\rm m}#3\mbox{$^{\rm s}\mskip-7.6mu.\,$}#4$} %% \dechms{04}{14}{12}{9198} = RA en formato 04^h 14^m 12^s .9198

\newcommand{\decdms}[4]{$#1^{\circ}#2'#3\mbox{$''\mskip-7.6mu.\,$}#4$} %% \decdms{28}{12}{12}{199} = Dec en formato 28^o 12' 12'' .199

\newcommand{\decms}[3]{$#1'#2\mbox{$''\mskip-7.6mu.\,$}#3$} %% \decms{12}{12}{199} = Dec en formato 12' 12'' .199
\newcommand{\msec}[2]{$#1\mbox{$''\mskip-7.6mu.\,$}#2$}

\newcommand{\rams}[3]{$#1^{\rm m}#2\mbox{$^{\rm s}\mskip-7.6mu.\,$}#3$}

\newcommand{\gaia}{{{\it Gaia}~DR2} }

\shorttitle{Astrometry of Orion radio stars}
\shortauthors{Dzib et al.}

\begin{document}

\title{A VLBA Survey of radio stars in the Orion Nebula Cluster: II. Astrometry}

\correspondingauthor{Sergio A. Dzib}
\email{sdzib@mpifr-bonn.mpg.de}

\author[0000-0001-6010-6200]{Sergio A. Dzib}
\affiliation{Max-Planck-Institut f\"ur Radioastronomie, Auf dem H\"ugel 69,
 D-53121 Bonn, Germany}
\author[0000-0001-8694-4966]{Jan Forbrich}
\affiliation{Centre for Astrophysics Research, University of Hertfordshire,
College Lane, Hatfield AL10 9AB, UK}
\affiliation{Center for Astrophysics $\vert$ Harvard \& Smithsonian, 60 Garden St, MS 72, Cambridge, MA 02138, USA}
\author[0000-0001-7223-754X]{Mark J. Reid}
\affiliation{Center for Astrophysics $\vert$ Harvard \& Smithsonian, 60 Garden St, MS 72, Cambridge, MA 02138, USA}
\author[0000-0001-6459-0669]{Karl M. Menten}
\affiliation{Max-Planck-Institut f\"ur Radioastronomie, Auf dem H\"ugel 69,
 D-53121 Bonn, Germany}

\begin{abstract}
From Very Long Baseline Array (VLBA) observations we  previously identified a population 
of 123 young stellar systems with nonthermal radio emission toward the core of the Orion 
Nebula Cluster (ONC).  We find optical sources in the {\it Gaia} DR2 catalog 
for 34 of them within 0\rlap{.}$''$2 of the radio positions. Most of the radio
sources  are likely {\it companions} of {\it Gaia} detections. However, there 
are  11 stars whose VLBA position differ from {\it Gaia} by $<1.6$~mas, and
some of these radio sources probably are the direct counterparts of the 
optical stars. We  are able to obtain radio proper motions for  23
stars. Combining the stellar proper motions derived 
from  the VLBA and \gaia data we find the global motion and velocity dispersion 
of the ONC to be 
($\mu_\alpha*,\mu_\delta$)=($1.20\pm0.09,0.18\pm0.09$)~mas~yr$^{-1}$ and
($\sigma_{\mu_\alpha*},\sigma_{\mu_\delta}$)=($0.84\pm0.09,1.30\pm0.09$)~mas~yr$^{-1}$.
Finally, we looked for ordered motions by estimating the means of scalar and 
vectorial products, which results in 
$\mathbf{\overline{v\cdot\hat{r}}}=-0.61\pm1.00$~km~s$^{-1}$ and 
$\mathbf{\overline{v\times\hat{r}}}=0.57\pm0.95$~km~s$^{-1}$. These do not show 
indications that the young stellar cluster is in expansion, contraction or 
rotation. 
\end{abstract}


\nopagebreak 
\section{Introduction}\label{intro}

The Orion Nebula Cluster (ONC), at a distance of about 
400~pc \citep{menten2007,kim2008,kounkel2017}, is the nearest
region in which massive stars have formed ${<2}$~million years 
ago \citep{muench2008}. It contains a rich stellar population
composed of a few high-mass stars, which already are on the 
main sequence, and many low-mass pre-main sequence   
young stellar objects 
(YSOs). With its $\sim 3500$ members, the ONC  is the most 
extensively studied young, partially embedded, star cluster
{\citep[e.g., ][]{hillenbrand1997,getman2005,daRio2012}.} 

The Kleinmann-Low (KL) Nebula is located a fraction 
of a parsec behind the ONC \citep{zuckerman1973,genzel1989}. 
Within it, the deeply embedded Becklin-Neugebauer object (BN) 
and a number of other embedded near infrared (NIR) sources 
 are evidence for more recent star formation.

At radio wavelengths, a rich population of compact sources has 
been found, many of which represent young stars in the ONC that 
are visible at NIR and optical wavelengths whereas others are 
associated with optically invisible NIR sources in the KL region. 
In the radio and the NIR range, the most prominent radio source 
in this region is BN 
\citep{garay87,chu87,zapata2004,kounkel2014,forbrich2016}. 
Some very deeply embedded sources show no optical or even infrared 
counterparts at all, but are still detected at X-ray and/or radio 
wavelengths \citep[see, e.g., ][]{mentenreid1995,forbrich2008}. Multi-epoch 
high-resolution interferometric studies of these YSOs at radio 
wavelengths have helped to constrain the ONC distance 
\citep{sandstom2007,hirota2007,menten2007,kounkel2017}, study 
their kinematics \citep[e.g.][]{dzib2017} and uncover sources 
related to the dynamical decay, some 500 { yr} ago, of a multiple stellar system within
the KL Nebula \citep[e.g.,][and references 
therein]{gomez2008,rodriguez2017,rodriguez2020}.

Magnetically active young low-mass stars produce non-thermal 
radio emission from their coronae \citep{feigelson1999}. The
coronae have sizes of at most a few stellar radii \citep{gudel2002}, and
have high brightness temperatures ($>10^6\,$K). Thus, they
provide excellent targets for observations with the Very Long 
Baseline Interferometry technique (VLBI) which provides 
astrometric precision of the order of tens of micro-arcseconds 
\citep[][]{reid2014}.

In order to characterize the non-thermal population of the ONC 
and study its kinematics, we initiated a campaign of high angular 
resolution observations with the Very Long Baseline Array (VLBA) 
targeting all 557 compact radio sources\footnote{We use the 
nomenclature [FRM2016] followed by the catalog number in 
\citet{forbrich2016} to name these sources.} known to exist in 
the region \citep[e.g.,][]{forbrich2016, dzib2017}. In a 
companion paper we focused on the detection criteria and sample 
definition (Forbrich et al. submitted; hereafter Paper I), and 
in future articles we shall discuss in detail the properties the 
whole sample. In this paper we report the measured positions and 
proper motions for sources detected in two or more epochs.

\begin{table}%{!ht}
\small
\begin{center}
\renewcommand{\arraystretch}{0.9}
\caption{Observed epochs.}
\begin{tabular}{ccccccc}\hline\hline
      & Date &   Synthesized beam & \\
Epoch & yyyy/mm/dd & $(\theta_{\rm maj}\times\theta_{\rm min}; {\rm P.A.})$&\#$_{\rm Ant.}$\\
\hline
1 & 2015/10/26 &  \msec{0}{00471}$\times$\msec{0}{00165}; $+22.1^{\circ}$&8\\
2 & 2017/10/26  & \msec{0}{00435}$\times$\msec{0}{00135}; $-17.3^{\circ}$&7\\
3 & 2017/10/27 &  \msec{0}{00407}$\times$\msec{0}{00139}; $-17.8^{\circ}$&7\\
4 & 2018/10/26 &  \msec{0}{00283}$\times$\msec{0}{00116}; $-02.1^{\circ}$&10\\
\hline\hline
\label{tab:epochs}
\end{tabular}
\end{center}
\tablecomments{Columns are (left to right): Epoch of the observation, civil date, FWHM major and minor axis and position angle (E of N) of the synthesized beam and the number of antennas used.}
\end{table}

\section{Observations} \label{sect:observations}

A full description of the observations is given in Paper~I, and only a 
brief summary follows. Four C-band observations, centered at 7.196~GHz,
were carried out with the VLBA in three different years 
(Tab.~\ref{tab:epochs}). Three of these observations were arranged to 
be made at the same day of the year, October 26, in 2015, 2017, and 2018.
The observations in 2015 and 2017 resulted in larger synthesized beam 
sizes because the array's most outlying antennas (Mauna Kea or St. Croix), 
which provide the longest baselines, were not operational (see Paper~I for 
full details). An additional epoch was obtained on 2017 October 27. 
Given the low declination of the ONC, the synthesized beam 
is elongated in north-south direction, resulting in a better astrometric 
precision in right ascension than in the declination direction. 

The 557 known sources within the primary beam were correlated using the 
DiFX software correlator \citep{deller2011}, which generates separate visibility 
data sets for each source position. The phase calibrator was the quasar 
J0541$-$0541 correlated at the position R.A.=\rahms{05}{41}{38}{083384}; 
Dec.=\decdms{-05}{41}{49}{42839} i.e.,  with angular separations { from} the ONC 
of 1\rlap{.}$^\circ$6 in right ascension and 0\rlap{.}$^\circ$3 in declination. 
The correlated position was { 0.24~mas} off in right ascension { and 0.07~mas 
off in declination} from the most recent determined position of this quasar 
provided by the AstroGeo website\footnote{This website (\url{http://astrogeo.org/})
provides a catalog of accurate positions for $\sim$17000 extragalactic sources 
with compact radio emission. { The positions are updated four times per year. 
We have consulted the current catalog {\it rfc\_2020c} released in September 23rd, 
2020.} The position of J0541$-$0541 is listed as 
R.A.=rahms{05}{41}{38}{083368}$\pm0\rlap{.}^{\rm s}000007$; 
Dec.=decdms{-05}{41}{49}{42846}$\pm0\rlap{.}''00011$.}, and source positions 
in this paper have been corrected to reflect the new calibrator position.

Data calibration was performed using the Astronomical Image Processing System 
software  \citep[AIPS;][]{Greisen_2003}. The calibration was done following 
standard procedures and images of detected sources were produced with pixel 
sizes of 50~$\mu$as.  See Paper I for further information on the image
processing and a detailed description on detection criteria. Images are exported 
in FITS format and read into the CASA software. General properties of the images 
are listed in Table~\ref{tab:epochs}. Source positions are measured with the 
CASA task {\it imfit}. The formal position errors derived from {\it imfit} range 
from 0.02 to 0.60~mas. However, we note that at this frequency, systematic residual 
position errors are expected to be $\approx0.1$ mas per degree-of-separation 
between the phase calibrator and the target \citep{reid2017},  however it is not 
clear if this systematic residual is direction dependent. 
In our case, we have to consider a total systematic error of 0.16~mas appropriate
for the 1\rlap{.}$^\circ$6 separation between the phase calibrator and the ONC. 
The 0.16~mas error affects both coordinates, { and} for each coordinate we
consider ${0.16~{\rm mas}/\sqrt{2}}\simeq\,$0.11~mas{. This} value was 
added in quadrature to the formal position errors derived from {\it imfit}.

\startlongtable
\begin{deluxetable*}{crrrrrcccc}
\fontsize{11}{9}\selectfont
\renewcommand{\arraystretch}{0.88}
%\tablenum{1}
\tablecaption{Positions of detected sources in the VLBA images and their formal errors as derived from {\it imfit}. %Full Table 2 will be only available online as a machine--readable table.
\label{tab:posF}}
\tablewidth{0pt}
\small
\tablehead{
\colhead{[FRM} & \colhead{} & 
\colhead{$\alpha_{{\rm J}2000}$} & 
\colhead{$\sigma_\alpha$} &
\colhead{$\delta_{{\rm J}2000}$} & 
\colhead{$\sigma_\delta$} & %_{\rm err}$} 
\colhead{}&
\colhead{$\Delta\theta$}&\colhead{$\Delta\alpha$}&\colhead{$\Delta\delta$}\\ 
\colhead{2016]}& \colhead{Epoch} & \colhead{$5^{\rm h}$} & \colhead{$\mu$s} &\colhead{$-5^{\circ}$}  & \colhead{$\mu$as} &\colhead{{\it Gaia} ID}&\colhead{(mas)}&\colhead{(mas)}&\colhead{(mas)}
}
%\decimalcolnumbers
\startdata
{ 2-1} & 3  &  \rams{34}{55}{975458} &  10  &  \decms{23}{13}{02414}  &  { 330}  & 3017364613086735360 & $22.4\pm0.3$ & $16.1\pm0.3$ & $15.6\pm0.3$ \\
{ 2-2}&  4  &  \rams{34}{55}{974448} &  1  &  \decms{23}{13}{03823}  &  44 & { 3017364613086735360} & ${1.5\pm0.3}$ & ${-0.6\pm0.3}$ & ${1.3\pm0.3}$\\
10 &  4  &  \rams{35}{06}{283539} &  1  &  \decms{22}{02}{66564}  &  114 & 3017364303848915072 & $6.6\pm1.0$ & $2.3\pm1.2$ & $6.1\pm1.0$ \\  
11 &  3  &  \rams{35}{06}{416885} &  5  &  \decms{24}{21}{34747}  &  200 &  \nodata &  \nodata &  \nodata &  \nodata\\
14 &  2  &  \rams{35}{07}{243891} &  7  &  \decms{22}{26}{30111}  &  324 &  \nodata &  \nodata &  \nodata &  \nodata\\
18 &  1  &  \rams{35}{09}{675491} &  5  &  \decms{23}{55}{91216}  &  203 &  \nodata &  \nodata &  \nodata &  \nodata\\ 
 &  {2}  &  \rams{35}{09}{676642} &  10  &  \decms{23}{55}{93895}  &  { 480} &  \nodata &  \nodata &  \nodata &  \nodata\\ 
21 &  2+3  &  \rams{35}{09}{769705} &  5  &  \decms{23}{26}{88897}  &  341 & 3017363994611276032 & $1.0\pm0.4$ & $0.1\pm0.3$ & $0.9\pm0.4$ \\
22 &  3  &  \rams{35}{09}{769946} & 3 &  \decms{21}{28}{34796} & 168 & 3017364406917802752 & $5.7\pm1.6$ & $-0.6\pm1.7$ & $5.4\pm1.6$\\ 
24 &  1  &  \rams{35}{09}{882645} &  10  &  \decms{23}{38}{33131}  &  468 &  \nodata &  \nodata &  \nodata &  \nodata\\
25 &  2+3  &  \rams{35}{10}{043939} &  7  &  \decms{21}{21}{93669}  &  414 &  \nodata &  \nodata &  \nodata &  \nodata\\
30 &  2  &  \rams{35}{10}{252267} &  10  &  \decms{21}{57}{11309}  &  301 & 3017364406926998784\tablenotemark{a} & $148.8\pm0.8$ & $-148.6\pm0.8$ & $7.0\pm1.1$ \\
32 &  2+3  &  \rams{35}{10}{494684} &  10  &  \decms{22}{45}{75147}  &  361 & 3017364235129434624 & $26.7\pm0.6$ & $1.1\pm0.6$ & $-26.7\pm0.6$\\
35 &  2+3  &  \rams{35}{10}{597692} &  7  &  \decms{22}{55}{66427}  &  294 &  \nodata &  \nodata &  \nodata &  \nodata\\
37 &  3  &  \rams{35}{10}{736683} &  6  &  \decms{23}{44}{72569}  &  278 & 3017363994611283840 & $56.8\pm0.4$ & $36.8\pm0.3$ & $-43.3\pm0.4$\\
42 &  4  &  \rams{35}{10}{940386} &  2  &  \decms{23}{26}{41294}  &  168 &  \nodata &  \nodata &  \nodata &  \nodata\\
47 & 2+3  &  \rams{35}{11}{255552} &  12  &  \decms{22}{16}{80621}  &  543 &  \nodata &  \nodata &  \nodata &  \nodata\\
53 &  4  &  \rams{35}{11}{562674} &  2  &  \decms{24}{48}{09454}  &  138 &  \nodata &  \nodata &  \nodata &  \nodata\\ 
55 &  3  &  \rams{35}{11}{615114} &  5  &  \decms{20}{22}{21699}  &  225 &  \nodata &  \nodata &  \nodata &  \nodata\\ 
64 &  3  &  \rams{35}{11}{725033} &  7  &  \decms{25}{12}{78408}  &  385 &  \nodata &  \nodata &  \nodata &  \nodata\\
66 &  1  &  \rams{35}{11}{804276} & 3 &  \decms{21}{49}{26414} & 129 &  \nodata &  \nodata &  \nodata &  \nodata\\ 
 &  2  &  \rams{35}{11}{804447} & 1 &  \decms{21}{49}{26663} & 74 &  \nodata &  \nodata &  \nodata &  \nodata\\ 
 &  3  &  \rams{35}{11}{804440} & 1 &  \decms{21}{49}{26656} & 78 &  \nodata &  \nodata &  \nodata &  \nodata\\ 
 &  4  &  \rams{35}{11}{804558} & 1 &  \decms{21}{49}{26779} & 48 &  \nodata &  \nodata &  \nodata &  \nodata\\
70 &  3  &  \rams{35}{11}{955534} &  7  &  \decms{20}{32}{36771}  &  280 &  \nodata &  \nodata &  \nodata &  \nodata\\ 
72 &  3  &  \rams{35}{12}{049462} &  8  &  \decms{22}{12}{07822}  &  315 &  \nodata &  \nodata &  \nodata &  \nodata\\
75 &  4  &  \rams{35}{12}{141226} &  1  &  \decms{24}{33}{46621}  &  106 &  \nodata &  \nodata &  \nodata &  \nodata\\
86 &  2+3  &  \rams{35}{12}{600988} &  7  &  \decms{21}{45}{50568}  &  265 &  \nodata &  \nodata &  \nodata &  \nodata\\ 
93 &  1  &  \rams{35}{12}{847280} & 7 &  \decms{21}{33}{97869} & 244 &  \nodata &  \nodata &  \nodata &  \nodata\\ 
98 &  1  &  \rams{35}{12}{964607} &  6  &  \decms{23}{54}{70582}  &  204 &  \nodata &  \nodata &  \nodata &  \nodata\\
122 &  3  &  \rams{35}{13}{428370} &  14  &  \decms{22}{52}{27556}  &  228 &  \nodata &  \nodata &  \nodata &  \nodata\\
127  &  3  &  \rams{35}{13}{506072} &  12  &  \decms{22}{19}{94996}  &  360 &  \nodata &  \nodata &  \nodata &  \nodata\\
129 &  3  &  \rams{35}{13}{529464} &  5  &  \decms{21}{12}{75039}  &  231 &  \nodata &  \nodata &  \nodata &  \nodata\\ 
130 &  1  &  \rams{35}{13}{586150} & 13 &  \decms{23}{55}{26292} & 456 &  \nodata &  \nodata &  \nodata &  \nodata\\ 
 &  2  &  \rams{35}{13}{586319} & 2 &  \decms{23}{55}{26712} & 111 &  \nodata &  \nodata &  \nodata &  \nodata\\ 
 &  3  &  \rams{35}{13}{586308} & 1 &  \decms{23}{55}{26701} & 60 &  \nodata &  \nodata &  \nodata &  \nodata\\ 
 &  4  &  \rams{35}{13}{586438} & 1 &  \decms{23}{55}{26861} & 33 &  \nodata &  \nodata &  \nodata &  \nodata\\
133  &  3  &  \rams{35}{13}{646517} &  6  &  \decms{24}{09}{10777}  &  295 &  \nodata &  \nodata &  \nodata &  \nodata\\ 
135 &  2  &  \rams{35}{13}{702626} &  5  &  \decms{21}{49}{17879}  &  260 &  \nodata &  \nodata &  \nodata &  \nodata\\
137 &  3  &  \rams{35}{13}{708093} &  5  &  \decms{25}{08}{17479}  &  265 &  \nodata &  \nodata &  \nodata &  \nodata\\
&  { 4}  &  \rams{35}{13}{708136} &  { 5}  &  \decms{25}{08}{21691}  &  { 314} &  \nodata &  \nodata &  \nodata &  \nodata\\
148 &  2  &  \rams{35}{13}{957145} &  11  &  \decms{23}{20}{47154}  &  501 &  \nodata &  \nodata &  \nodata &  \nodata\\ 
149 &  2+3  &  \rams{35}{13}{903353} &  5  &  \decms{24}{09}{28667}  &  278 &  \nodata &  \nodata &  \nodata &  \nodata\\
154 &  1  &  \rams{35}{13}{972262} & 4 &  \decms{24}{09}{84011} & 149 &  \nodata &  \nodata &  \nodata &  \nodata\\ 
 &  3  &  \rams{35}{13}{972311} & 7 &  \decms{24}{09}{83995} & 310 &  \nodata &  \nodata &  \nodata &  \nodata\\ 
 &  4  &  \rams{35}{13}{972375} & 1 &  \decms{24}{09}{83914} & 40 &  \nodata &  \nodata &  \nodata &  \nodata\\ 
158 &  2  &  \rams{35}{14}{054152} & 11 &  \decms{23}{38}{45757} & 428 & 3017363960251335424 & $10.7\pm0.8$ & $-7.1\pm0.8$ & $-7.9\pm0.8$\\ 
 &  4  &  \rams{35}{14}{054327} & 4 &  \decms{23}{38}{45545} & 320 &  \nodata &  \nodata &  \nodata &  \nodata\\ 
161 &  1  &  \rams{35}{14}{058854} &  33  &  \decms{20}{12}{573445}  &  518 &  \nodata &  \nodata &  \nodata &  \nodata\\
167 &  4  &  \rams{35}{14}{196172} &  3  &  \decms{26}{21}{14518}  &  241 &  \nodata &  \nodata &  \nodata &  \nodata\\
170 &  4  &  \rams{35}{14}{262925} &  4  &  \decms{22}{35}{45441}  &  196 &  \nodata &  \nodata &  \nodata &  \nodata\\ 
176 &  4  &  \rams{35}{14}{339783} &  3  &  \decms{21}{17}{44629}  &  219 &  \nodata &  \nodata &  \nodata &  \nodata\\
177-1 &  1  &  \rams{35}{14}{335495} & 9 &  \decms{23}{17}{42271} & 373 &  \nodata &  \nodata &  \nodata &  \nodata\\ 
 &  4  &  \rams{35}{14}{336155} & 2 &  \decms{23}{17}{42214} & 216 &  \nodata &  \nodata &  \nodata &  \nodata\\ 
\,177-2  &  4  &  \rams{35}{14}{335835} &  1  &  \decms{23}{17}{39939}  &  56 &  \nodata &  \nodata &  \nodata &  \nodata\\ 
182 &  3  &  \rams{35}{14}{424573} &  9  &  \decms{21}{26}{70387}  &  344 &  \nodata &  \nodata &  \nodata &  \nodata\\ 
184 &  1  &  \rams{35}{14}{501740} & 14 &  \decms{22}{38}{69711} & 459 &  \nodata &  \nodata &  \nodata &  \nodata\\ 
 &  2  &  \rams{35}{14}{501784} & 4 &  \decms{22}{38}{70306} & 212 &  \nodata &  \nodata &  \nodata &  \nodata\\ 
  &  { 3}  &  \rams{35}{14}{501787} & { 6} &  \decms{22}{38}{70296} & { 264} &  \nodata &  \nodata &  \nodata &  \nodata\\ 
 &  { 4}  &  \rams{35}{14}{501883} & {1} &  \decms{22}{38}{70192} & { 52} &  \nodata &  \nodata &  \nodata &  \nodata\\ 
188 &  2  &  \rams{35}{14}{505056} &  11  &  \decms{23}{10}{34936}  &  403 &  \nodata &  \nodata &  \nodata &  \nodata\\
189 &  1  &  \rams{35}{14}{545411} &  9  &  \decms{23}{15}{99164}  &  267 &  \nodata &  \nodata &  \nodata &  \nodata\\
196 &  1  &  \rams{35}{14}{655810} &  4  &  \decms{22}{33}{74156}  &  152 & 3017364132048871168 & $17.9\pm0.2$ & $-16.7\pm0.2$ & $6.5\pm0.2$\\
197 &  2  &  \rams{35}{14}{646950} &  5  &  \decms{20}{42}{23801}  &  260 & 3017365918756031744 & $173.8\pm2.1$ & $-165.6\pm2.1$ & $52.7\pm1.8$\\
198 &  3  &  \rams{35}{14}{665959} & 10 &  \decms{22}{11}{28275} & 348 &  \nodata &  \nodata &  \nodata &  \nodata\\ 
 &  4  &  \rams{35}{14}{665154} & 3 &  \decms{22}{11}{28841} & 215 &  \nodata &  \nodata &  \nodata &  \nodata\\
%201 &  4  &  \rams{35}{14}{665141} &  8  &  \decms{22}{11}{28887}  &  281 &  \nodata &  \nodata &  \nodata &  \nodata\\ 
203 &  2+3  &  \rams{35}{14}{731189} &  4  &  \decms{22}{29}{82403}  &  296 &  \nodata &  \nodata &  \nodata &  \nodata\\
205 &  1  &  \rams{35}{14}{794851} &  20  &  \decms{21}{53}{89877}  &  497 &  \nodata &  \nodata &  \nodata &  \nodata\\
211 &  1  &  \rams{35}{14}{898525} & 11 &  \decms{22}{25}{40704} & 380 &  \nodata &  \nodata &  \nodata &  \nodata\\ 
 &  2  &  \rams{35}{14}{898318} & 3 &  \decms{22}{25}{41142} & 153 &  \nodata &  \nodata &  \nodata &  \nodata\\ 
 &  3  &  \rams{35}{14}{898317} & 2 &  \decms{22}{25}{41168} & 134 &  \nodata &  \nodata &  \nodata &  \nodata\\ 
 &  4  &  \rams{35}{14}{898408} & 3 &  \decms{22}{25}{41948} & 280 &  \nodata &  \nodata &  \nodata &  \nodata\\ 
212 &  1  &  \rams{35}{14}{916016} &  2  &  \decms{22}{39}{20577}  &  82 & 3017364127743283584 & $25.6\pm0.3$ & $-25.3\pm0.3$ & $4.2\pm0.2$\\ 
222 &  3  &  \rams{35}{15}{234278} &  12  &  \decms{22}{56}{71107}  &  356 &  \nodata &  \nodata &  \nodata &  \nodata\\ 
227 &  3  &  \rams{35}{15}{340230} &  8  &  \decms{22}{18}{22981}  &  361 &  \nodata &  \nodata &  \nodata &  \nodata\\
230 &  4  &  \rams{35}{15}{393924} &  6  &  \decms{22}{33}{11715}  &  149 &  \nodata &  \nodata &  \nodata &  \nodata\\
232 &  4  &  \rams{35}{15}{391542} &  4  &  \decms{22}{29}{88937}  &  214 &  \nodata &  \nodata &  \nodata &  \nodata\\ 
240 &  1  &  \rams{35}{15}{521395} &  5  &  \decms{23}{37}{49236}  &  271 &  \nodata &  \nodata &  \nodata &  \nodata\\ 
241 &  1  &  \rams{35}{15}{555065} & 4 &  \decms{25}{14}{12514} & 144 & 3017360966647238272 & $1.3\pm0.2$ & $-0.2\pm0.2$ & $1.3\pm0.2$\\ 
 &  2  &  \rams{35}{15}{555280} & 2 &  \decms{25}{14}{12257} & 130 &  \nodata &  \nodata &  \nodata &  \nodata\\ 
 &  3\tablenotemark{b}  &  \rams{35}{15}{555262} & 13 &  \decms{25}{14}{12239} & 444 &  \nodata &  \nodata &  \nodata &  \nodata\\ 
 &  4  &  \rams{35}{15}{555413} & 1 &  \decms{25}{14}{12075} & 47 &  \nodata &  \nodata &  \nodata &  \nodata\\ 
242 &  2+3  &  \rams{35}{15}{588076} &  12  &  \decms{21}{26}{87028}  &  382 &  \nodata &  \nodata &  \nodata &  \nodata\\
249 &  4  &  \rams{35}{15}{749637} &  6  &  \decms{23}{38}{74543}  &  302 &  \nodata &  \nodata &  \nodata &  \nodata\\
250 &  1  &  \rams{35}{15}{773727} & 2 &  \decms{23}{09}{87026} & 81 & 3017364127743288704 & $1.0\pm0.2$ & $-0.3\pm0.2$ & $1.0\pm0.2$\\ 
 &  2  &  \rams{35}{15}{773899} & 1 &  \decms{23}{09}{86841} & 86 &  \nodata &  \nodata &  \nodata &  \nodata\\ 
 &  3  &  \rams{35}{15}{773902} & 1 &  \decms{23}{09}{86840} & 65 &  \nodata &  \nodata &  \nodata &  \nodata\\ 
 &  4  &  \rams{35}{15}{773989} & 1 &  \decms{23}{09}{86683} & 39 &  \nodata &  \nodata &  \nodata &  \nodata\\ 
254 &  1  &  \rams{35}{15}{829235} & 2 &  \decms{23}{14}{15060} & 72 & 3017364132050194688 & $183.8\pm0.2$ & $33.8\pm0.2$ & $180.6\pm0.2$\\ 
 &  2  &  \rams{35}{15}{829898} & 1 &  \decms{23}{14}{15607} & 67 &  \nodata &  \nodata &  \nodata &  \nodata\\ 
 &  3  &  \rams{35}{15}{829878} & 1 &  \decms{23}{14}{15598} & 72 &  \nodata &  \nodata &  \nodata &  \nodata\\ 
 &  4  &  \rams{35}{15}{830213} & 1 &  \decms{23}{14}{15819} & 34 &  \nodata &  \nodata &  \nodata &  \nodata\\ 
273 &  4  &  \rams{35}{16}{096551} &  2  &  \decms{23}{27}{94239}  &  171 &  \nodata &  \nodata &  \nodata &  \nodata\\ 
285 &  2+3  &  \rams{35}{16}{184889} &  3  &  \decms{21}{32}{84282}  &  192 &  \nodata &  \nodata &  \nodata &  \nodata\\ 
300 &  1  &  \rams{35}{16}{356075} &  5  &  \decms{24}{02}{82463}  &  309 &  \nodata &  \nodata &  \nodata &  \nodata\\ 
303 &  3  &  \rams{35}{16}{411129} &  8  &  \decms{22}{12}{37498}  &  406 &  \nodata &  \nodata &  \nodata &  \nodata\\ 
314 &  1  &  \rams{35}{16}{642910} &  8  &  \decms{20}{26}{63953}  &  330 &  \nodata &  \nodata &  \nodata &  \nodata\\ 
319 &  1  &  \rams{35}{16}{766695} & 7 &  \decms{24}{04}{25260} & 284 & 3017363955944478976 & $12.7\pm0.4$ & $-4.3\pm0.4$ & $-12.0\pm0.4$\\ 
 &  2  &  \rams{35}{16}{766633} & 7 &  \decms{24}{04}{24963} & 341 &  \nodata &  \nodata &  \nodata &  \nodata\\ 
 &  3  &  \rams{35}{16}{766632} & 10 &  \decms{24}{04}{24969} & 389 &  \nodata &  \nodata &  \nodata &  \nodata\\ 
 &  4  &  \rams{35}{16}{766638} & 3 &  \decms{24}{04}{24756} & 254 &  \nodata &  \nodata &  \nodata &  \nodata\\ 
321 &  2  &  \rams{35}{16}{738295} &  17  &  \decms{23}{28}{30463}  &  446 &  \nodata &  \nodata &  \nodata &  \nodata\\ 
326 &  3  &  \rams{35}{16}{935631} &  4  &  \decms{22}{10}{22981}  &  225 &  \nodata &  \nodata &  \nodata &  \nodata\\ 
327 &  1  &  \rams{35}{16}{983560} &  15  &  \decms{23}{33}{02930}  &  409 &  \nodata &  \nodata &  \nodata &  \nodata\\ 
335 &  2  &  \rams{35}{17}{042397} &  10  &  \decms{23}{39}{63166}  &  416 &  \nodata &  \nodata &  \nodata &  \nodata\\
339 &  4  &  \rams{35}{17}{121826} &  8  &  \decms{24}{34}{50190}  &  238 &  \nodata &  \nodata &  \nodata &  \nodata\\
343 &  1  &  \rams{35}{17}{220600} &  15  &  \decms{21}{31}{70144}  &  423 & 3017364368261463936 & $11.3\pm0.4$ & $3.3\pm0.3$ & $10.8\pm0.4$\\ 
347 &  4  &  \rams{35}{17}{334877} &  1  &  \decms{22}{36}{12379}  &  154 &  \nodata &  \nodata &  \nodata &  \nodata\\
350 &  1  &  \rams{35}{17}{392308} & 11 &  \decms{22}{03}{62402} & 414 &  \nodata &  \nodata &  \nodata &  \nodata\\ 
 &  2\tablenotemark{b}  &  \rams{35}{17}{392548} & 9 &  \decms{22}{03}{62441} & 410 &  \nodata &  \nodata &  \nodata &  \nodata\\ 
 &  4  &  \rams{35}{17}{392764} & 1 &  \decms{22}{03}{62460} & 146 &  \nodata &  \nodata &  \nodata &  \nodata\\ 
357 &  2+3  &  \rams{35}{17}{503387} &  23  &  \decms{21}{06}{03158}  &  575 &  \nodata &  \nodata &  \nodata &  \nodata\\ 
360 &  {\bf 3}  &  \rams{35}{17}{529123} &  13  &  \decms{21}{45}{79922}  &  665 &  \nodata &  \nodata &  \nodata &  \nodata\\ 
364 &  3  &  \rams{35}{17}{677673} &  6  &  \decms{23}{41}{15188}  &  244 &  \nodata &  \nodata &  \nodata &  \nodata\\
367 &  2  &  \rams{35}{17}{710690} &  5  &  \decms{24}{43}{20288}  &  322 &  \nodata &  \nodata &  \nodata &  \nodata\\
373 &  2+3  &  \rams{35}{17}{869362} &  16  &  \decms{22}{15}{27217}  &  868 &  \nodata &  \nodata &  \nodata &  \nodata\\
378 &  1  &  \rams{35}{17}{952561} & 2 &  \decms{22}{45}{43436} & 80 & 3017364127743299328 & $1.0\pm0.2$ & $-0.5\pm0.2$ & $0.8\pm0.2$\\ 
 &  2  &  \rams{35}{17}{953051} & 1 &  \decms{22}{45}{42909} & 64 &  \nodata &  \nodata &  \nodata &  \nodata\\ 
 &  3  &  \rams{35}{17}{953037} & 1 &  \decms{22}{45}{42902} & 67 &  \nodata &  \nodata &  \nodata &  \nodata\\ 
 &  4  &  \rams{35}{17}{953269} & 1 &  \decms{22}{45}{42740} & 33 &  \nodata &  \nodata &  \nodata &  \nodata\\ 
382 &  1\tablenotemark{b}  &  \rams{35}{18}{030668} & 12 &  \decms{22}{05}{39440} & 605 & 3017364166397351296\tablenotemark{a} & ${1.6\pm0.8}$ & ${1.1\pm1.0}$ & ${1.1\pm0.8}$\\ 
 &  4  &  \rams{35}{18}{031363} & 1 &  \decms{22}{05}{40002} & 140 &  \nodata &  \nodata &  \nodata &  \nodata\\ 
389 &  1  &  \rams{35}{18}{216537} &  7  &  \decms{23}{36}{05593}  &  304 & 3017364063330467072 & $194.0\pm0.3$ & $123.7\pm0.2$ & $-149.5\pm0.3$\\ 
398 &  4  &  \rams{35}{18}{305932} &  1  &  \decms{25}{05}{71411}  &  109 &  \nodata &  \nodata &  \nodata &  \nodata\\
400 &  1  &  \rams{35}{18}{372944} & 2 &  \decms{22}{37}{42754} & 74 & 3017364162103039104 & $1.1\pm0.2$ & $-0.4\pm0.2$ & $1.0\pm0.2$\\ 
 &  2  &  \rams{35}{18}{373244} & 1 &  \decms{22}{37}{42642} & 72 &  \nodata &  \nodata &  \nodata &  \nodata\\ 
 &  3  &  \rams{35}{18}{373239} & 1 &  \decms{22}{37}{42623} & 63 &  \nodata &  \nodata &  \nodata &  \nodata\\ 
 &  4  &  \rams{35}{18}{373427} & 1 &  \decms{22}{37}{42542} & 32 &  \nodata &  \nodata &  \nodata &  \nodata\\ 
402 &  4  &  \rams{35}{18}{388443} &  2  &  \decms{20}{20}{34921}  &  202 & 3017365948809434624 & $4.3\pm2.0$ & $3.1\pm2.4$ & $-0.6\pm2.7$\\
408 &  3  &  \rams{35}{18}{550783} &  12  &  \decms{21}{27}{88721}  &  288 &  \nodata &  \nodata &  \nodata &  \nodata\\
414-1 &  1  &  \rams{35}{18}{660441} & 4 &  \decms{20}{33}{83380} & 184 &  \nodata &  \nodata &  \nodata &  \nodata\\ 
 &  2  &  \rams{35}{18}{660697} & 14 &  \decms{20}{33}{83502} & 862 &  \nodata &  \nodata &  \nodata &  \nodata\\ 
 &  4  &  \rams{35}{18}{660855} & 4 &  \decms{20}{33}{83662} & 257 &  \nodata &  \nodata &  \nodata &  \nodata\\ 
414-2 &  3  &  \rams{35}{18}{673470} & 3 &  \decms{20}{33}{71209} & 187 & 3017365880089958912 & $11.6\pm1.1$ & $8.6\pm1.2$ & $7.7\pm1.1$\\ 
 &  4  &  \rams{35}{18}{673562} & 1 &  \decms{20}{33}{71145} & 97 &  \nodata &  \nodata &  \nodata &  \nodata\\ 
426 &  3  &  \rams{35}{18}{979086} &  7  &  \decms{25}{08}{14860}  &  305 &  \nodata &  \nodata &  \nodata &  \nodata\\
435  &  3  &  \rams{35}{19}{213224} &  4  &  \decms{22}{50}{68184}  &  183 & 3017364097690179712 & $1.1\pm0.4$ & $0.5\pm0.6$ & $0.8\pm0.4$\\ 
440 &  4  &  \rams{35}{19}{493670} &  3  &  \decms{22}{58}{86720}  &  204 &  \nodata &  \nodata &  \nodata &  \nodata\\ 
456 &  3  &  \rams{35}{19}{855122} &  8  &  \decms{23}{57}{55121}  &  385 &  \nodata &  \nodata &  \nodata &  \nodata\\ 
457 &  2  &  \rams{35}{20}{067807} &  16  &  \decms{21}{06}{22523}  &  485 &  \nodata &  \nodata &  \nodata &  \nodata\\
459 &  1  &  \rams{35}{20}{135592} &  3  &  \decms{21}{33}{63052}  &  114 & 3017365643877358464 & $1.5\pm0.2$ & $-0.3\pm0.2$ & $1.5\pm0.2$\\ 
462 &  4  &  \rams{35}{20}{168294} &  1  &  \decms{26}{39}{08568}  &  94 & 3017360313812222720 & $1.0\pm0.3$ & $-0.3\pm0.4$ & $0.9\pm0.3$ \\ 
466  &  2  &  \rams{35}{20}{226344} &  3  &  \decms{20}{56}{81211}  &  204 & 3017365884396345600 & $165.2\pm3.7$ & $42.7\pm4.0$ & $159.5\pm3.7$\\ 
  &  3  &  \rams{35}{20}{226344} &  1  &  \decms{20}{56}{81220}  &  68 &  \nodata &  \nodata &  \nodata &  \nodata\\ 
467 &  3  &  \rams{35}{20}{296851} &  5  &  \decms{25}{04}{46652}  &  195 &  \nodata &  \nodata &  \nodata &  \nodata\\ 
468 &  4  &  \rams{35}{20}{400552} &  2  &  \decms{22}{13}{62575}  &  160 & 3017364162103048704 & $4.6\pm0.6$ & $4.1\pm0.6$ & $1.9\pm0.5$\\ 
470 &  4  &  \rams{35}{20}{481920} &  31  &  \decms{24}{20}{28599}  &  851 &  \nodata &  \nodata &  \nodata &  \nodata\\
477 &  4  &  \rams{35}{20}{665562} &  13  &  \decms{22}{45}{41051}  &  659 &  \nodata &  \nodata &  \nodata &  \nodata\\
480 &  1 &  \rams{35}{20}{725215} & 11 &  \decms{21}{44}{33936} & 396 & 3017365643879095040 & $36.1\pm0.3$ & $21.3\pm0.2$ & $29.2\pm0.4$\\ 
 &  4  &  \rams{35}{20}{724666} & 1 &  \decms{21}{44}{33392} & 97 &  \nodata &  \nodata &  \nodata &  \nodata\\ 
485 &  1  &  \rams{35}{21}{049414} & 3 &  \decms{23}{49}{00528} & 132 & 3017361108393637120 & $0.5\pm0.2$ & $-0.4\pm0.2$ & $0.3\pm0.2$\\ 
 &  2  &  \rams{35}{21}{049475} & 5 &  \decms{23}{49}{00468} & 259 &  \nodata &  \nodata &  \nodata &  \nodata\\ 
 &  3  &  \rams{35}{21}{049468} & 2 &  \decms{23}{49}{00431} & 103 &  \nodata &  \nodata &  \nodata &  \nodata\\ 
 &  4  &  \rams{35}{21}{049575} & 1 &  \decms{23}{49}{00162} & 88 &  \nodata &  \nodata &  \nodata &  \nodata\\ 
501 &  2  &  \rams{35}{22}{116650} &  8  &  \decms{24}{32}{31589}  &  311 &  \nodata &  \nodata &  \nodata &  \nodata\\
508 &  4  &  \rams{35}{22}{299739} &  2  &  \decms{24}{14}{14305}  &  171 &  \nodata &  \nodata &  \nodata &  \nodata\\
509 &  4  &  \rams{35}{22}{364446} &  3  &  \decms{25}{01}{99665}  &  330 &  \nodata &  \nodata &  \nodata &  \nodata\\
512 &  3  &  \rams{35}{22}{676241} &  6  &  \decms{23}{06}{17739}  &  227 &  \nodata &  \nodata &  \nodata &  \nodata\\ 
514 &  3  &  \rams{35}{22}{824118} &  4  &  \decms{25}{47}{67696}  &  200 & 3017360730428788608 & $199.8\pm1.5$ & $-194.3\pm1.5$ & $-46.6\pm1.2$\\
515 &  4  &  \rams{35}{22}{877839} &  4  &  \decms{24}{57}{59939}  &  273 &  \nodata &  \nodata &  \nodata &  \nodata\\
520 &  4  &  \rams{35}{23}{495180} &  9  &  \decms{20}{01}{67517}  &  406 & 3017365983169196416\tablenotemark{a} & $8.9\pm4.6$ & $4.3\pm6.3$ & $1.6\pm6.3$\\ 
521 &  1  &  \rams{35}{23}{596407} &  19  &  \decms{25}{26}{71826}  &  594 &  \nodata &  \nodata &  \nodata &  \nodata\\ 
522 &  3  &  \rams{35}{23}{680511} &  8  &  \decms{23}{46}{19317}  &  360 &  \nodata &  \nodata &  \nodata &  \nodata\\ 
525 &  2  &  \rams{35}{23}{953379} &  7  &  \decms{25}{09}{49805}  &  261 &  \nodata &  \nodata &  \nodata &  \nodata\\ 
526 &  1  &  \rams{35}{24}{016722} &  19  &  \decms{23}{14}{09561}  &  579 &  \nodata &  \nodata &  \nodata &  \nodata\\ 
527 &  3  &  \rams{35}{24}{273278} &  7  &  \decms{25}{18}{86354}  &  268 &  \nodata &  \nodata &  \nodata &  \nodata\\ 
530 &  4  &  \rams{35}{24}{468753} &  3  &  \decms{24}{00}{91711}  &  179 &  \nodata &  \nodata &  \nodata &  \nodata\\ 
534 &  3  &  \rams{35}{24}{853747} &  4  &  \decms{21}{00}{81741}  &  169 &  \nodata &  \nodata &  \nodata &  \nodata\\ 
535 &  4  &  \rams{35}{25}{015058} &  4  &  \decms{24}{38}{53626}  &  205 &  \nodata &  \nodata &  \nodata &  \nodata\\
537 &  2  &  \rams{35}{25}{088440} &  4  &  \decms{23}{46}{79491}  &  204 & 3017361074033139200 & $10.2\pm0.3$ & $8.9\pm0.3$ & $-5.0\pm0.4$\\ 
547 &  3  &  \rams{35}{26}{397825} &  2  &  \decms{25}{00}{72346}  &  142 & 3017360833515044480 & $95.5\pm0.4$ & $-66.8\pm0.4$ & $68.3\pm0.4$\\ 
552 &  3  &  \rams{35}{29}{588773} &  8  &  \decms{23}{12}{23421}  &  276 &  \nodata &  \nodata &  \nodata &  \nodata\\ 
555 &  2  &  \rams{35}{31}{445661} &  8  &  \decms{25}{16}{43957}  &  480 &  \nodata &  \nodata &  \nodata &  \nodata\\
557 &  3  &  \rams{35}{14}{950741} & 7 &  \decms{23}{39}{24439} & 324 & 3017363960237919616 & $15.1\pm0.8$ & $9.1\pm0.9$ & $-12.0\pm0.8$\\ 
 &  4  &  \rams{35}{14}{950932} & 5 &  \decms{23}{39}{24586} & 424 &  \nodata &  \nodata &  \nodata &  \nodata\\ 
\enddata
\tablecomments{
%Table 2 is published in its entirety in the machine-readable format.
%      A portion is shown here for guidance regarding its form and content. 
Columns are (left to right): Source number [from \citet{forbrich2016}],
      epoch, J2000 right ascension and declination, both with uncertainties, % Columns are (left to right): Source number from \citep{forbrich2016}
      %epoch, 
      {\it Gaia} ID, the total separation 
%($\Delta\theta$) 
between the Gaia and VLBA positions and 
the separation in both coordinate directions. The separations are defined as: $\Delta\alpha=(\alpha_{\rm VLBA}-\alpha_{\rm \gaia})\cdot\cos{\delta}$, $\Delta\delta=(\delta_{\rm VLBA}-\delta_{\rm \gaia})$ and $\Delta\theta=\sqrt{\Delta\alpha^2 + \Delta\delta^2}$.}
      \tablenotetext{a}{Source in \gaia catalog, no proper motion and parallaxes are given. The angular separations from the radio source are estimated using the positions in epoch 2015.5 as given in the \gaia archive.  A Gaia ID source number is also listed when there is coincidence 
%within $0\rlap{.}''2$ from the radio source. 
      }
     \tablenotetext{b}{ Source detected at a level between 6.0 -- 6.5 times the image's noise.
      }
\end{deluxetable*}
$ $

\begin{figure*}%[!th]
   \centering
  \includegraphics[height=0.74\textwidth, trim=0 0 0 0, clip]{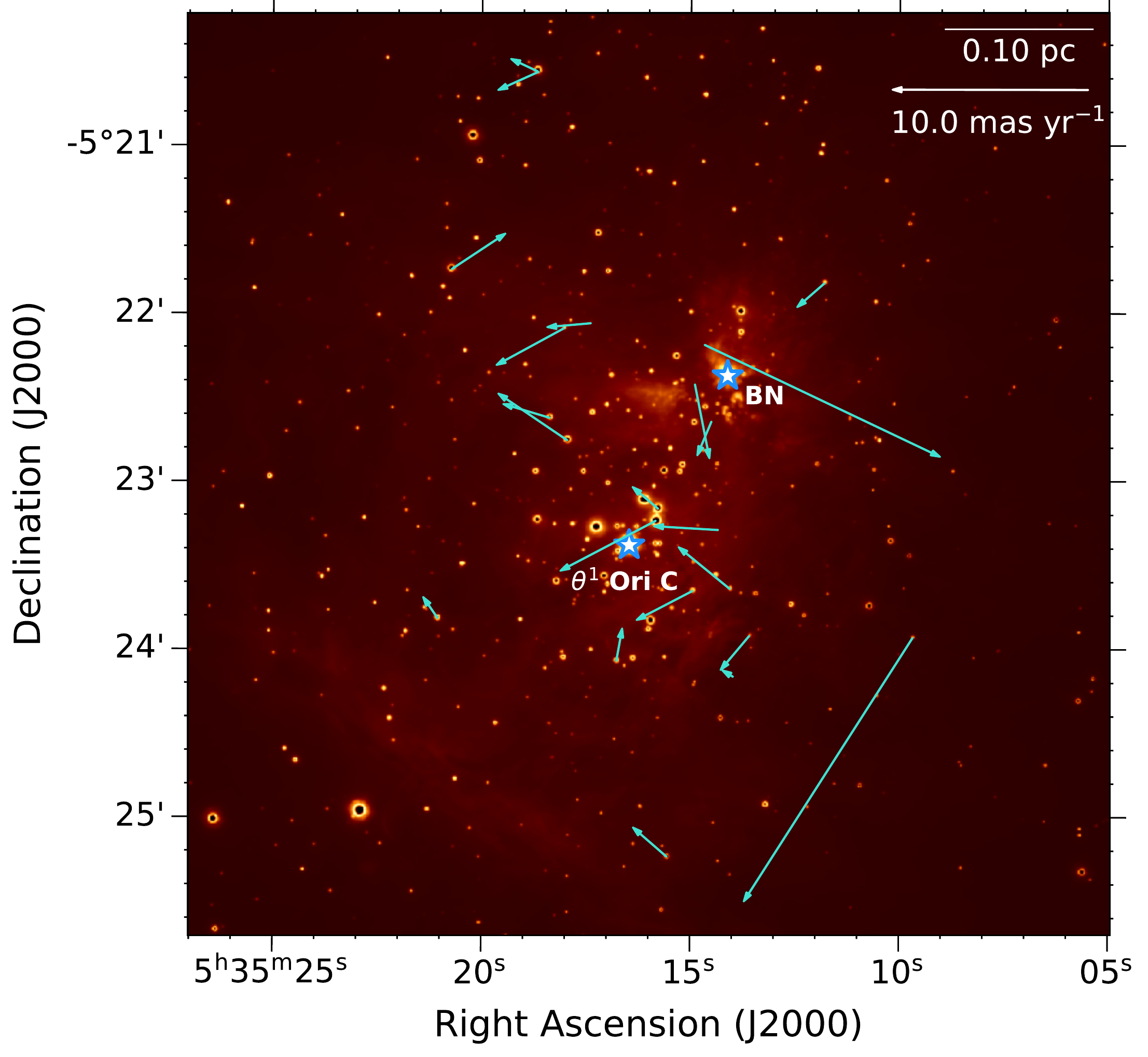}
   \caption{Background: VISION  2.1\,$\mu$m image of the ONC \citep{meingast2016}. 
  Cyan arrows indicate the proper motion vectors of YSOs in the ONC-Trapezium region from Table~\ref{tab:RAM},  with the exception of source [FRM2016]~137 because of its very large nominal proper motion (see text). The location of the two massive stars $\theta^1$~Ori~C and BN are indicated with blue stars.}
   \label{fig:PM}
\end{figure*}

\begin{deluxetable*}{cccrrcrrc}
\tablecaption{Sources with determined proper motions from VLBA observations.} 
\tablehead{
\colhead{}     &\colhead{}       &\colhead{} &\multicolumn{2}{c}{VLBA} &\colhead{}  & \multicolumn{2}{c}{\gaia\ }&\colhead{}\\ \cline{4-5}\cline{7-8}
\colhead{}    & \colhead{Other} & \colhead{Spectral}&  \multicolumn{1}{c}{$\mu_\alpha^{*}$} & \multicolumn{1}{c}{$\mu_\delta$} & \colhead{$\Delta\theta$} &  \multicolumn{1}{c}{$\mu_\alpha^{*}$} & \multicolumn{1}{c}{$\mu_\delta$}&\colhead{M$_{\rm low}$}\\
\colhead{$[$FRM2016$]$} & \colhead{Name} & \colhead{Type\tablenotemark{\footnotesize a}} &\multicolumn{1}{c}{(mas yr$^{-1}$)}  &  \multicolumn{1}{c}{(mas yr$^{-1}$)}&(mas)&\multicolumn{1}{c}{(mas yr$^{-1}$)}  &  \multicolumn{1}{c}{(mas yr$^{-1}$)}&\colhead{(M$_\odot$)}\\
\colhead{(1)} & \colhead{(2)} & \colhead{(3)}&\multicolumn{1}{c}{(4)} & \multicolumn{1}{c}{(5)} & (6) & \multicolumn{1}{c}{(7)} & \multicolumn{1}{c}{(8)} & \colhead{(9)}
%\hline
}
\startdata
\enskip{ 18\tablenotemark{\footnotesize $\dagger$}}& { COUP 338} & \nodata  &${8.59\pm0.11}$ &${-13.40\pm0.28}$&\nodata& \nodata & \nodata & \nodata \\ 
66& GMR A &\nodata & $1.38\pm0.10$ & $-1.21\pm0.07$&\nodata&\nodata&\nodata&\nodata \\
130& COUP 554 &\nodata &$1.43\pm0.19$ & $-1.76\pm0.15$&\nodata&\nodata&\nodata&\nodata \\
137& \nodata &\nodata &${0.63\pm0.19}$ & ${-42.12\pm0.44}$&\nodata&\nodata&\nodata&\nodata \\
154& COUP 594 &\nodata&$0.54\pm0.14$ & $0.32\pm0.13$&\nodata&\nodata&\nodata&\nodata\\
\enskip158\tablenotemark{\footnotesize $\dagger$}& COUP 602 &M3&$2.62\pm0.23$ & $2.12\pm0.55$&$10.7\pm0.8$&$3.23\pm 0.32$ &$ 1.25\pm 0.27$&0.02\\ %M3
\quad\,177-1\tablenotemark{\footnotesize $\dagger$}& COUP 625 &\nodata&$3.28\pm0.07$ & $0.19\pm0.15$&\nodata&\nodata&\nodata &\nodata\\
\enskip184 & GMR H &\nodata&${0.71\pm0.28}$ & ${-1.71\pm0.91}$&\nodata&\nodata&\nodata&\nodata \\ 
\enskip198\tablenotemark{\footnotesize $\dagger$} & COUP 647 & ORBS &$-12.05\pm0.21$& $-5.67\pm0.43$&\nodata&\nodata&\nodata&\nodata\\ 
211 & GMR D &\nodata &$-0.76\pm0.63$& $-3.76\pm1.24$&\nodata&\nodata&\nodata&\nodata\\
241 & V1501 Ori&K4-M1&$1.70\pm0.12$ & $1.46\pm0.09$&$1.3\pm0.2$& $1.22\pm 0.13$ &$ 1.53\pm 0.11$&\nodata\\ %K4-M1
250 &$\theta^1$ Ori E&B5-B8, G0-G5&$1.31\pm0.05$ & $1.11\pm0.14$&$1.0\pm0.2$&$1.61\pm 0.12$ &$ 1.23\pm 0.11$&\nodata\\ %B5-B8, G0-G5   
254 &$\theta^1$ Ori A$_2$&B1.5&$4.87\pm0.07$ & $-2.56\pm0.12$&$183.8\pm0.2$&$1.54\pm 0.16$ &$ 0.12\pm 0.14$&5.6\\ %O9-B1.5
319 & V1279 Ori&K2-K6&$-0.31\pm0.12$ & $1.66\pm0.13$&$12.7\pm0.4$&$1.92\pm 0.52$ &$ 2.61\pm 0.43$&0.12\\ %K2-K6 
350 & COUP 874 &\nodata &$2.23\pm0.36$ & $-0.20\pm0.15$&\nodata&\nodata&\nodata&\nodata\\  
378 & GMR G & G-K3&$3.54\pm0.08$ & $2.38\pm0.21$&$1.0\pm0.2$&$3.77\pm 0.10$ &$ 2.34\pm 0.08$&\nodata\\ %G-K3
\enskip382\tablenotemark{\footnotesize $\dagger$} & COUP 942 & G-M2&$3.46\pm0.08$ & $-1.88\pm0.21$&${1.6\pm0.8}$\tablenotemark{\footnotesize b}&\nodata&\nodata&\nodata\\ 
400 & GMR F &G8-M2&$2.38\pm0.13$ & $0.70\pm0.08$&$1.1\pm0.1$&$2.13\pm 0.10$& $0.78\pm 0.08$&\nodata\\   %G8-M2
\quad414-1& \nodata&\nodata &$2.05\pm0.08$ & $-0.93\pm0.11$&\nodata&\nodata&\nodata&\nodata\\ 
\enskip\quad414-2\tablenotemark{\footnotesize $\dagger$}& COUP 985&F8-K4 &$1.38\pm0.15$ & $0.64\pm0.25$&$11.6\pm1.2$&$-1.45\pm 0.52$& $-1.31\pm 0.43$&\nodata\\ %F8-K4 
\enskip480\tablenotemark{\footnotesize $\dagger$} & V1230 Ori& B1&$-2.74\pm0.07$& $1.81\pm0.14$&$36.1\pm0.4$&$2.60\pm 0.11$& $-1.72\pm 0.10$&2.4\\   %B
485 & GMR V & G8-K5&$0.74\pm0.24$ & $1.07\pm0.52$&$0.4\pm0.2$& $0.05\pm 0.12$& $-1.01\pm 0.11$&\nodata\\  %G8-K5
\enskip557\tablenotemark{\footnotesize $\dagger$} & COUP 672&K5-M2&$2.85\pm0.20$&$-1.48\pm0.55$&$15.1\pm0.8$&$0.86\pm 0.37$ &$0.55\pm 0.30$&0.2 \\ %K5-M2
\enddata
\label{tab:RAM}
\tablecomments{Columns are (left to right): Source number from \citet{forbrich2016},
identification names from other surveys, spectral types from \cite{hillenbrand2013}, proper motions in right ascension and declination, both with uncertainties, total angular separation from \gaia sources, as given in Table 2, and their corresponding proper motions, and the mass lower limit estimated from the differences between VLBA and \gaia\ proper motions, as discussed in the text.}
\tablenotetext{ \dagger }{Source detected only in two epochs. The errors may be underestimated.}
\tablenotetext{a}{Spectral types as reported by \citet{hillenbrand2013}, for multiple systems these do not necessarily represent the counterparts of the radio sources.}
\tablenotetext{b}{Source in \gaia catalog, no proper motion and parallaxes are given. The angular separations from the radio source are estimated using the positions in epoch 2015.5 as given in the \gaia archive.}
\end{deluxetable*}

\section{Results}

In Paper I, we described the criteria used for source detection. A total of 123 
stellar systems were detected with brightness levels ${>6.5}$ times the noise. 
Most of the VLBA detections  match this criterion in only one epoch.
 However, once the source has been clearly detected in at least two epochs, we have lowered the detection criterion to be ${>6}$ times the noise for the remaining epochs if the position is in line with the clear detections. In one case, [FRM2016]~382, with just a single VLBA detection at $>$6.5 times the noise level, we also used a threshold of 6.0 times the noise since its position is consistent with the position of a star in the \gaia\ catalog.

The measured source positions 
are listed in Table~\ref{tab:posF}. In a small number of cases we detected 
multiple sources related to a single source found in the lower resolution VLA 
observations.

\subsection{Detection of multiple sources}

In the images of [FRM2016]~414 only one source is detected in epochs 1, 2 and 3. 
However, at epoch 4, two compact radio sources with a separation, $\Delta\theta$, of 0\rlap{.}$''$22 are 
 clearly detected. Given this large separation, compared with the angular resolution, 
we confirm that the radio sources detected at epochs 1 and 2 are associated with 
the eastern source in epoch 4, while the source at epoch 3 is associated with the
western source in epoch 4.   We refer to the source detected in three epochs as 
[FRM2016]~414-1 and the source detected in two epochs as [FRM2016]~414-2.

Similarly, source [FRM2016]~177 appears as a single source in epoch~1, while in 
epoch~4 two sources are clearly detected with a separation of 23.2~mas. A visual 
inspection of the images allows us to associate the southern source at { epoch}~4 
with the source detected at epoch~1.  We refer to southern and northern sources
in epoch 4 as [FRM2016]~177-1 and [FRM2016]~177-2, respectively.

Two VLBA detections are related to the VLA source [FRM2016]~2; both were single 
detections in epochs 3 and 4. The source detected in epoch 4 is separated from 
the source in epoch 3 by 20.6~mas. As we will discuss later we discard the possibility of a fast 
moving source since the last detection in epoch 4 is only  1.6~mas from a source 
in the \gaia\ catalog, which has a small proper motion of ${1.64\pm0.10}$~mas~yr$^{-1}$,
indicating that the radio detections  are from two different sources. 
As in the previous cases, we named these sources [FRM2016]~2-1 and [FRM2016]~2-2.

\subsection{Proper motions}

The observations were scheduled on nearly the same day of the year in order to 
remove the effects of parallax from estimates of proper motion. This allowed us 
to estimate motions with a minimum of two epochs separated by at least one year.
For those sources detected in three or four epochs, the motion fits produced a 
reduced $\chi^{2}$ near unity, indicating that our error estimates %were 
are realistic.
 Three exceptions, [FRM2016]~184, and 211, are discussed in Section 4.5, where we also 
discuss the three sources with proper motions $>10$~mas~yr$^{-1}$ ([FRM2016]~18, 137 and 198).
The fitted proper motions for  23 sources are listed in Table~\ref{tab:RAM}
and plottted in Fig.~\ref{fig:PM}.

%The particular case ONC184 need to be introduced and discussed.

\vspace{1cm}
\section{Discussion}

In this section, we compare our results with previous VLBA observations and to 
\gaia\ astrometry. The astrometric precision from our observations is similar 
to those achieved by {\it Gaia}. A comparison of positions and 
proper motion measurements from both telescopes can give us clues as to the nature 
of the detected sources, may show systematic differences between the results from different telescopes, and allow us to study the kinematics of the ONC.

It is interesting to note that the two instruments trace 
two different populations of stellar sources. {\it Gaia} measures stars that are
bright at optical wavelengths, have low extinction, and are not superimposed on bright nebulosity, while the VLBA 
observes (mostly)  magnetically active stars. Moreover, the VLBA targets are
not limited by the heavy dust obscuration of the BN-KL region.
Thus, the astrometry derived from VLBA and {\it Gaia} data is complementary 
when studying the kinematics of the ONC
and the BN/KL region close to it.

\subsection{Comparison with \gaia\ results: Positions}

We searched the \gaia\ catalog \citep{GC2018} for optical sources within 
0\rlap{.}$''$2 of the position of radio sources and found 34 such cases. 
The search radius was chosen considering the possibility of tight multiple 
components, unidentified in previous observations with lower angular resolution.
In order to determine the number of chance alignments between both catalogs
within this separation, first we consider that the primary beam of the VLBA 
observations covers an area of $\sim10^5$ arcsec$^2$, %where 
in which  we have detected 126 
radio sources. The total solid angle covered by our counterpart search is 
${126\times(0\rlap{.}''2)^2\times\pi\simeq15.8}$~arcsec$^2$. In the area covered 
by the VLBA primary beam there are 608 optical sources in the \gaia\ catalog. 
Thus, the number of \gaia\ sources projected onto our search area by chance is
$608\times(15.8/10^5)\simeq0.1$.

The positions from \gaia\ catalog are determined for epoch 2015.50 \citep{GC2018}. 
For a comparison with our VLBA data, the \gaia\ source positions and their errors 
have been extrapolated to the epoch of first detection of the corresponding radio 
source, including the effects of parallax.  In the case of the VLBA detections
related to source [FRM2016]~2 which are separated $20.6$~mas, we do the extrapolation 
of the {\it Gaia} source to both detected epochs and compared the positions. 
The position difference with the source detected in epoch 4 is significantly smaller
than for the source detected in epoch 3, indicating that the radio source detected 
in epoch 4 is related to the Gaia source. We will discuss source FRM2016]~2 
in more detail in Section~\ref{invsou}.  The separations of the radio from the 
\gaia\ sources are listed in Table~\ref{tab:posF}.

Since the position errors from both catalogs are below 1~mas, we expect the 
direct optical counterparts of radio sources to have radio versus optical separations 
of this magnitude. Only  11 of the 34 sources have separations less than 
1.6~mas, while 
the remaining  23 sources have $\Delta\theta\geq4.5$~mas. At the distance of the ONC, the radio emission from stellar coronae is expected to be unresolved by our VLBA 
observations. Even coronal flares and interactions between stellar coronae which can reach distances of tens of stellar radii \citep[e.g.,][]{gudel2002,massi2008} would 
still be unresolved with our observations. Thus, separations $\Delta\theta\geq4.5$~mas 
(or 1.8~AU at the distance of the ONC) may be an indication that these  23 radio sources 
trace a stellar companion to the star seen at optical wavelengths. A clear example is 
$\theta^1$~Ori~A. Here the primary component, a B0.5-type star, dominates the optical 
emission and
the strong radio source is 
a companion star, $\theta^1$~Ori~A$_2$ \citep{petr1998, petr2008, gravity2018}. 
\citet{gravity2018} recently measured the separation between both components 
in three epochs spanning over 1.13 years. A simple linear extrapolation of their measured relative positions 
to the epoch 2015.80 yields a total separation of $183.85\pm0.12$~mas,
which is consistent with the estimated separation
between the radio and the \gaia\ source of $\Delta\theta=183.8\pm0.2$~mas
(see also Table~\ref{tab:posF} and Sect. \ref{invsou} for a further discussion of this system).

\begin{figure}[!th]
   \centering
  \includegraphics[width=0.5\textwidth, trim=0 0 30 50, clip]{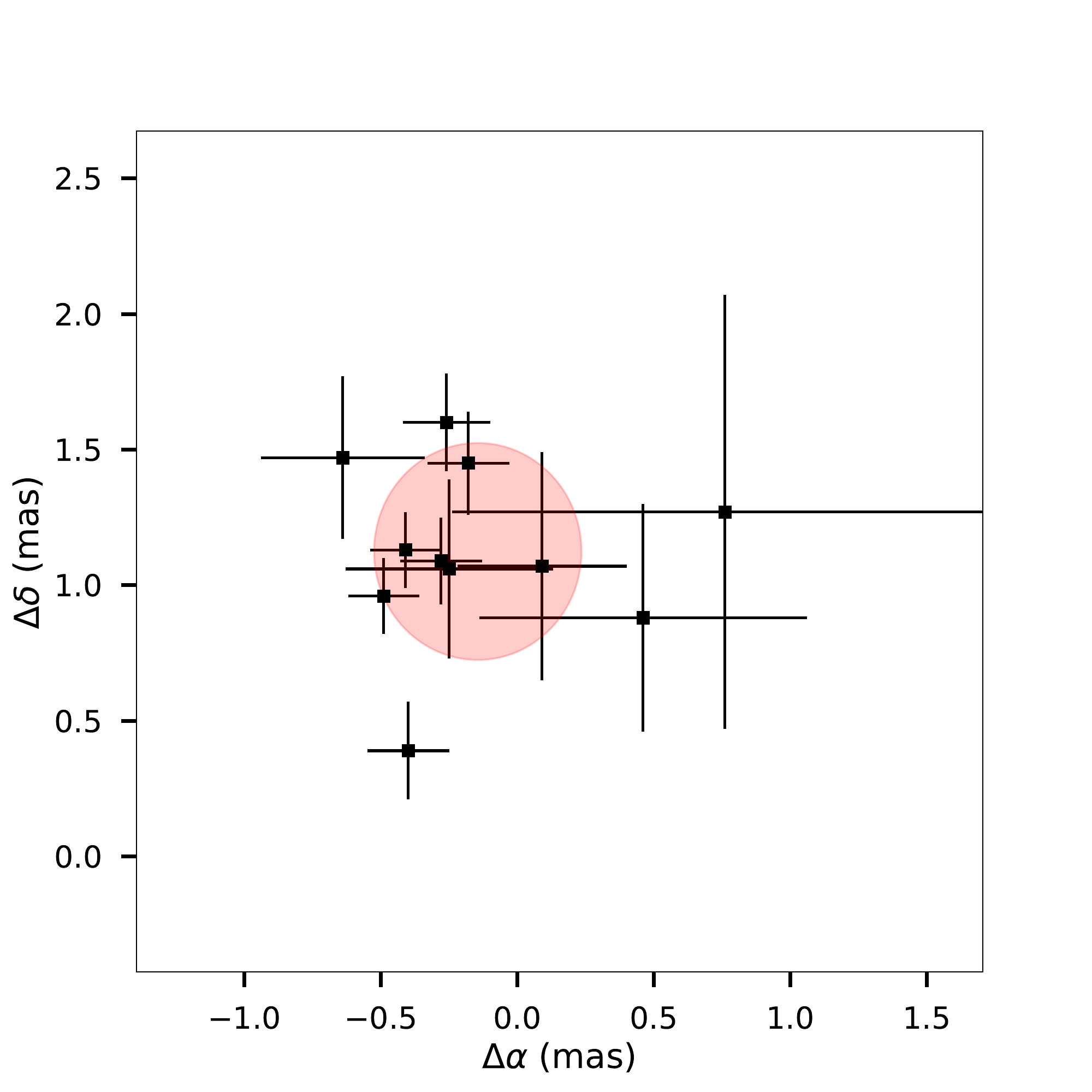}
   \caption{Comparison between VLBA  and {\it Gaia}  positions for the nine sources with total position separation ${<1.6}$~mas.  The pink ellipse is centered in mean separations and its size correspond to two times the error values.}
   \label{fig:VvsG_pos}
\end{figure}

The  11 radio sources with separations less than  1.6~mas could correspond to the true 
counterparts of the optical sources or to companions with separation $\leq0.6$~AU 
(1~mas$\equiv$0.4~AU). In Figure~\ref{fig:VvsG_pos}, we plot the distribution of the
separations of radio and optical sources in both coordinates. We find that the { weighted average 
of the separations} in right ascension and declination are 
$\overline{\Delta\alpha}={-0.32\pm0.15 ({\rm stat.}) \pm 0.16 ({\rm sys.})}$~mas and 
$\overline{\Delta\delta}={0.99\pm0.10 ({\rm stat.}) \pm 0.16 ({\rm sys.})}$~mas,
where the systematic uncertainty is associated with potential uncompensated ionospheric 
delays at the observing frequency of 7 GHz discussed in Section \ref{sect:observations}. { The statistical errors were estimated using the formalism of the standard errors of mean.}
Thus, while the mean difference between the radio and optical positions in right ascension 
is consistent with no shift between the two catalogs, the declination difference is 
statistically significant at the ${5\sigma}$ level.

In order to examine reference frame differences, we compile a list of quasars 
within a radius of $5^\circ$ of the ONC that have accurate positions determined by \gaia\ 
and the VLBA (see Appendix A).  We find that the mean position differences in right ascension 
and declination are ${-0.6\pm0.7}$~mas and ${0.0\pm0.6}$~mas, respectively. The observed systematic shift 
of ${1.03\pm0.16}$~mas for the ONC stars in declination is larger than that seen for the 
quasars; however, the uncertainty in the quasar positions precludes a decisive test at this time.
Similar studies, comparing VLBI and Gaia results, have been  carried out for stellar parallaxes and proper motions 
\citep[e.g.,][]{Xu2019}, also finding differences between both catalogs. To our knowledge 
this has not yet been done for stellar positions.  We conclude that the origin of the stellar 
position difference in declination between both catalogs remains uncertain.

\subsection{Comparison with \gaia\ results: Proper Motions}

Of the 23 radio sources for which we have determined proper motions, 
11 have a \gaia source within 0\rlap{.}$''$2,  and with estimated proper motions.
The \gaia\ proper motions and the separation between 
the radio source and the nearest optical counterpart 
are listed in Table~\ref{tab:RAM} columns (6) to (8). 
In some cases, the radio sources and their nearest
optical counterparts may not be directly associated as 
their separations are larger than  1.6~mas 
(see previous section).

\begin{figure*}[!th]
   \centering
  \includegraphics[height=0.38\textwidth, trim=0 0 0 0, clip]{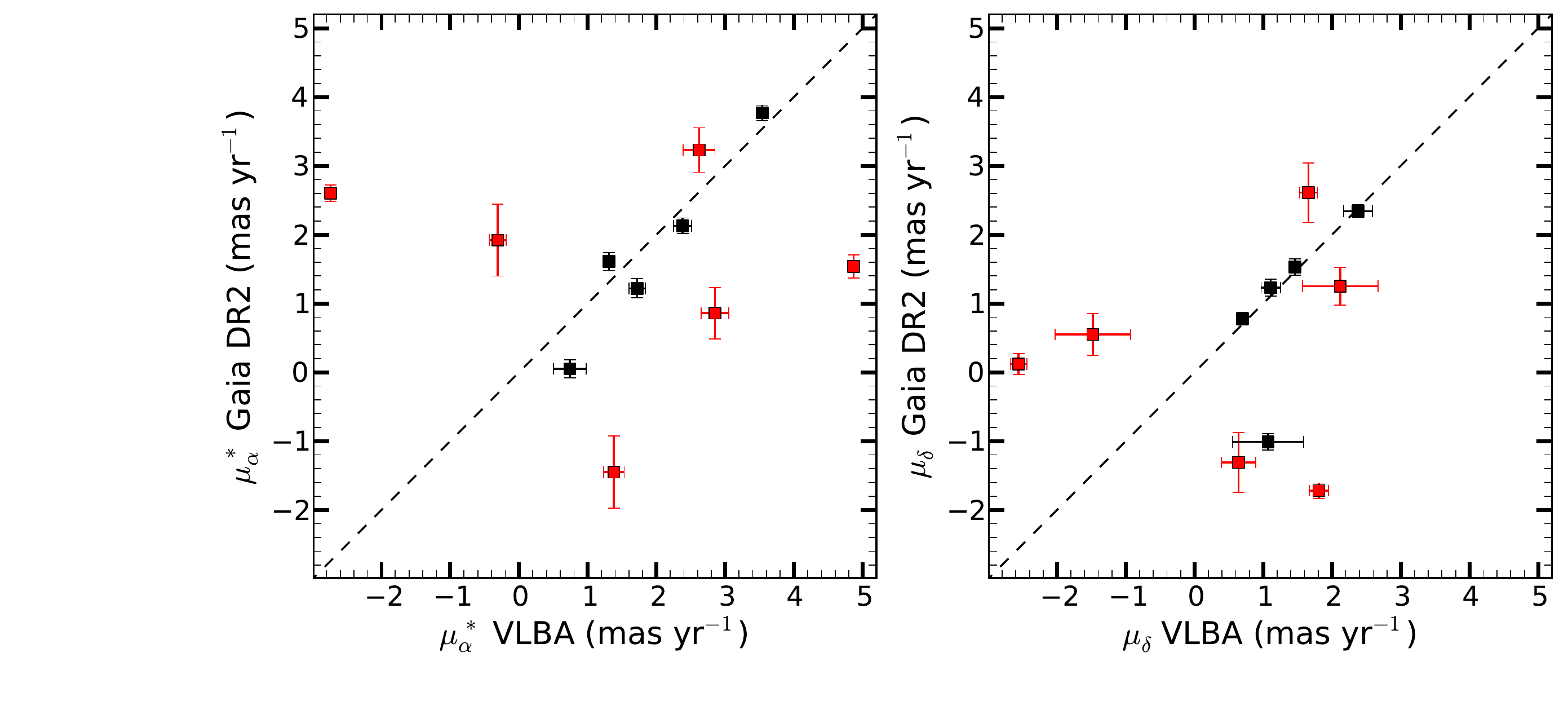}
   \caption{Comparison between VLBA  and {\it Gaia}  astrometry. {\it Left:} Proper motion 
   in right ascension. {\it Right:} Proper motions in declination. Black symbols 
   indicate sources where the difference in positions between VLBA  and Gaia DR2 
   is less than  1.6~mas, otherwise the symbols are red.}
   \label{fig:VvsG}
\end{figure*}

In Fig.~\ref{fig:VvsG} we plot the \gaia proper motion of the nearest optical source against
the VLBA proper motions. 
In this plot, black squares represent the sources  for which 
$\Delta\theta\leq1.6$~mas, and in red squares sources 
with  $\Delta\theta>1.6$~mas. 
There are  six sources with radio minus optical position differences 
which are larger than  1.6~mas, and for these we find that the  
proper motions are also inconsistent in both coordinates (see 
Table~\ref{tab:RAM} and Figure~\ref{fig:VvsG}).  In these cases,
the radio source may trace a stellar companion to the 
star seen at optical wavelengths, rather than its direct
counterpart. 

Assuming that these seven pairs of radio-optical sources 
are binary systems and that relative motion between both components
is dominated by the orbital motion, we may estimate a lower limit 
on the total mass of the system. Since the proper motions from 
\gaia\ and VLBA are not measured simultaneously, this analysis 
also implies that we are assuming that the orbital period 
of these candidate binaries is $\gg3$~years, 
the time separation between the epoch of reported parameters for \gaia\
and our last radio observation. 
The total difference in proper motion, $$\mu=\sqrt{(\mu_{{\alpha},{VLBA}}^*-\mu_{{\alpha},{Gaia}}^*)^2
+(\mu_{{\delta},{VLBA}}-\mu_{{\delta},{Gaia}})^2},$$ 
\noindent in these plausible binaries, can be 
used to estimate a lower limit on the mass of the system. 
The orbital velocity, for a circular orbit, is given by $$V=\sqrt{G\cdot M/a},$$
\noindent where $G$ is the gravitational constant, $M$
is the total mass of the system and ${a}$ is the distance between 
binary components.  The lower limit on mass is obtained assuming the 
orbit in the plane of the sky, then
$$M\geq a\cdot V^2/G$$
Lower limits for $V$ and $a$ can be obtained from 
the proper motion and the angular separation. 
At the ONC distance, these parameters are 
$V$[km~s$^1$]=$1.9\cdot\mu$[mas~yr$^{-1}$], and 
$a[km]=6\times10^{7}\cdot\Delta\theta$[mas]. 
The lower limits in the system masses are listed in Table~\ref{tab:RAM} column (9). 
Comparing these lower limits
with the spectral type, we see that systems with
early spectral types require larger lower limits, as would be expected
for the binary system hypothesis.

From Fig.~\ref{fig:VvsG} and Table~\ref{tab:RAM}, we find that for four of 
five stars with $\Delta\theta\leq1.6$~mas,   the differences of
the radio and optical proper motions are smaller than three times the quadrature
sum of {their} errors.
(the one outlier is source [FRM2016]~485~\citep[also known as GMR~V;][]{garay87}, whose VLBA and \gaia\ 
positions agrees within $0.4\pm0.2$~mas, but  the difference in proper
motion in declination is larger than five times the quadrature sum of errors).
The weighted average differences for these four stars 
are $-0.025\pm0.080$~mas~yr$^{-1}$ and 
$-0.070\pm0.090$~mas~yr$^{-1}$.  Adopting three times the errors for limits, 
these results indicate upper limits of $\sim$0.3~mas~yr$^{-1}$ 
for the differences in proper motions measured with VLBA and \gaia.
These limits will improve when more proper motions of radio 
sources can be determined.

\begin{figure}[!th]
   \centering
  \includegraphics[height=0.46\textwidth, trim=7 0 0 0, clip]{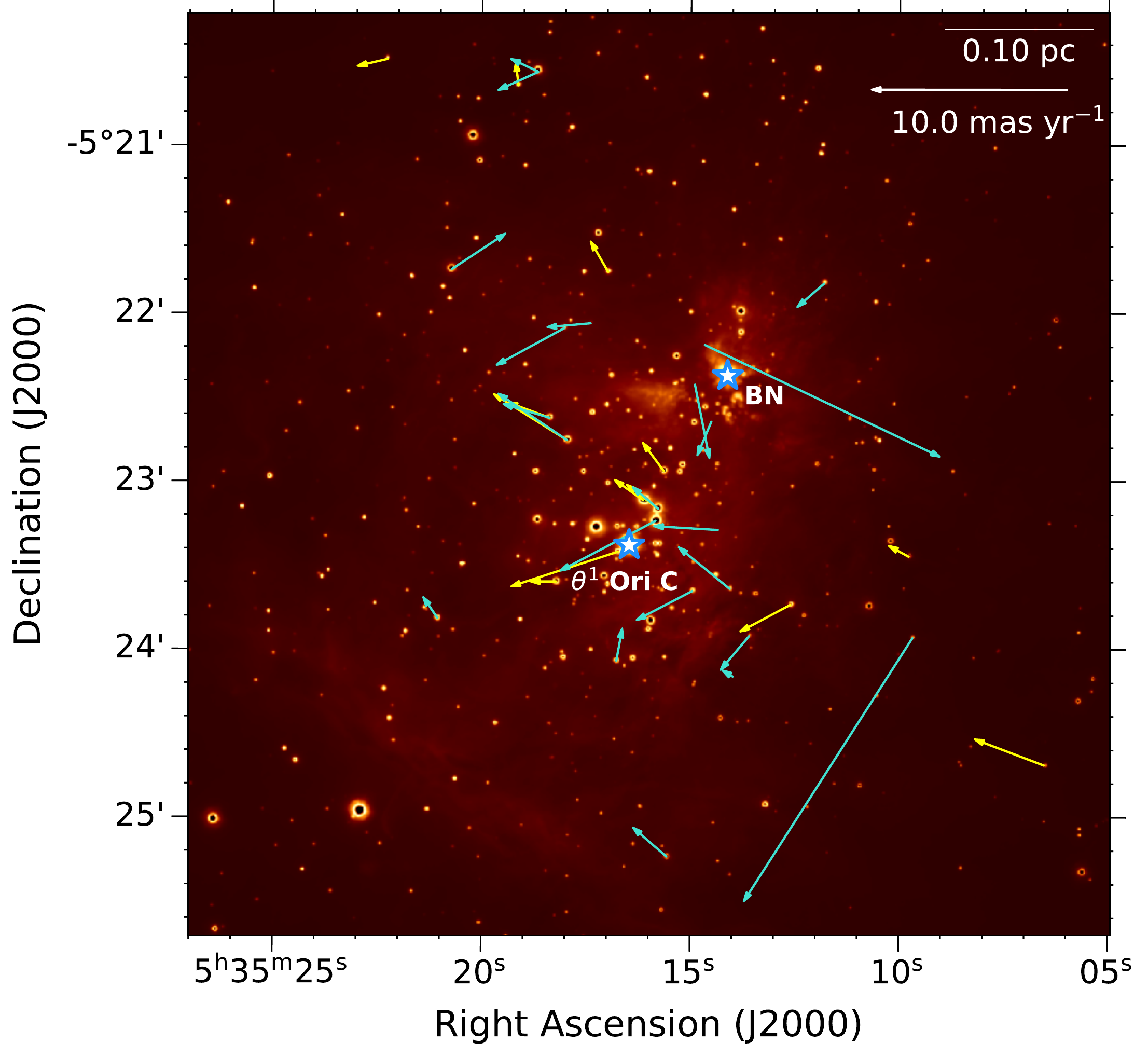}
   \caption{Same as Fig.~\ref{fig:PM}, but including in yellow arrows all the proper motions from Gaia DR2 catalog with RUWE$\leq$1.4 in the same region of our study.}
   \label{fig:VvsGPM}
\end{figure}

\begin{deluxetable*}{@{\extracolsep{2pt}}ccrrrrrr}
\tablecaption{Previous stellar proper motion (in mas yr$^{-1}$) measurements 
in the ONC with VLBA.}
\tablehead{
\colhead{}    & \colhead{Other} &\multicolumn{2}{c}{This work} & \multicolumn{2}{c}{\citet{menten2007}} & \multicolumn{2}{c}{\citet{kounkel2017}}  \\\cline{3-4}\cline{5-6}\cline{7-8}
\colhead{$[$FRM2016$]$}  &\colhead{name} & \multicolumn{1}{c}{$\mu_\alpha^{*}$} & \multicolumn{1}{c}{$\mu_\delta$} &\multicolumn{1}{c}{$\mu_\alpha^{*}$} & \multicolumn{1}{c}{$\mu_\delta$} &  \multicolumn{1}{c}{$\mu_\alpha^{*}$} & \multicolumn{1}{c}{$\mu_\delta$} 
}
\startdata
66 &GMR A &$1.38\pm0.10$ & $-1.21\pm0.07$
&$1.82\pm0.09$ & $-2.05\pm0.18$&$1.81\pm0.11$ & $-1.60\pm0.10$ \\
 184 &  GMR H &$0.71\pm0.28$ & $-1.71\pm0.91$
&\nodata &\nodata &$2.22\pm0.18$ & $-3.80\pm0.55$\\%K2017 3epochs 
250 & $\theta^{1}$\,Ori\,E &$1.31\pm0.05$&$1.11\pm0.14$
&\nodata &\nodata & $1.45\pm0.03$ & $1.02\pm0.08$\\   
254 & $\theta^{1}$\,Ori\,A$_2$ &$4.87\pm0.07$&$-2.56\pm0.12$
&$4.82\pm0.09$ &$-1.54\pm0.18$& $4.81\pm0.10$&$-2.53\pm0.12$\\ 
378 & GMR G &$3.54\pm0.08$ & $2.38\pm0.21$
&$4.29\pm0.17$ & $3.33\pm0.37$& $3.82\pm0.10$ & $1.60\pm0.17$\\ 
400 &GMR F&$2.38\pm0.13$ & $0.70\pm0.08$&$2.24\pm0.09$ & $0.66\pm0.18$& $2.38\pm0.08$ & $0.55\pm0.14$\\  
\enddata
\label{tab:ORAM}
\tablecomments{Columns are (left to right): Source number from \citet{forbrich2016},
identification names from other surveys, proper motions in right ascension and declination, both with uncertainties, from this work (see Table~\ref{tab:RAM}), \citet{menten2007}, and \citet{kounkel2017}.}
\end{deluxetable*}

\begin{deluxetable*}{llcccccccc}
\tablecaption{Kinematics of the Trapezium.}
\tablehead{
\colhead{}        & \colhead{$\overline{\mu_\alpha^{*}}$} & \colhead{$\overline{\mu_\delta}$} & \colhead{$\sigma_{\mu_{\alpha}}$}
& \colhead{$\sigma_{\mu_{\delta}}$} & \colhead{$\sigma_{v_{\alpha}}$\tablenotemark{\footnotesize a}}
& \colhead{$\sigma_{v_{\delta}}$\tablenotemark{\footnotesize a}} & \colhead{$\overline{\mathbf{v\cdot \hat{r}}}$} & \colhead{$\overline{\mathbf{v\times \hat{r}}}$}\\
\colhead{Data set} &  \colhead{(mas yr$^{-1}$)} &  \colhead{(mas yr$^{-1}$)} & \colhead{(mas yr$^{-1}$) }& 
 \colhead{(mas yr$^{-1}$)} &  \colhead{(km s$^{-1}$)} &  \colhead{(km s$^{-1}$)} & \colhead{(km s$^{-1}$)} & 
\colhead{(km s$^{-1}$)}
}%\\
\startdata
\hline
VLA\tablenotemark{\footnotesize b} &$1.0\pm0.1$& $-0.8\pm0.2$ & $1.1\pm0.1$ & $1.3\pm0.2$&$2.1\pm0.2$&
$2.5\pm0.4$ & $0.7\pm0.3$ & $-0.1\pm0.3$\\
 VLBA & $1.6\pm0.2$& $0.1\pm0.2$ & $1.4\pm0.2$ &$1.7\pm0.2$ 
& $2.7\pm0.4$ & $3.2\pm0.4$ & $-0.8\pm1.1$ & $0.5\pm1.1$\\
{\it Gaia} & $1.1\pm0.1$ & $\,\,0.2\pm0.1$ &$0.7\pm0.1$ &$1.2\pm0.1$ 
&$1.3\pm0.2$ & $2.2\pm0.2$ & $0.2\pm2.3$ & $0.8\pm1.9$\\ \hline
 VLBA+Gaia\tablenotemark{\footnotesize c} & $1.20\pm0.09$& $0.18\pm0.09$ & $0.84\pm0.09$ & $1.30\pm0.09$ 
&$1.60\pm0.17$ & $2.47\pm0.17$ & $-0.61\pm1.00$ & $0.57\pm0.95$\\ 
\hline\hline
\enddata
\label{tab:TK}
 \tablecomments{Columns are (left to right): Proper motions data set, mean of proper motions, proper motion dispersions, velocity dispersions (all these three parameter in right ascension and declination and with their uncertainties), mean of dot and cross products of the unit vector and velocity vectors, both with their uncertainties.}
\tablenotetext{a}{At a distance of 400 pc: 1.0~mas~yr$^{-1}\equiv1.9$~km~s$^{-1}$.}
\tablenotetext{b}{Values from \citet{dzib2017}.}
\tablenotetext{c}{Values are the variance weighted average from the indpendent results of VLBA and {\it Gaia} data sets.}
\end{deluxetable*}

\subsection{Comparison with previous VLBA results}

Previous VLBA proper motion results for young stars in the 
ONC were obtained by \citet{menten2007} and 
\citet{kounkel2017}. \citet{menten2007} observed four young 
stars at four epochs spanning 1.5 yr, while \citet{kounkel2017} obtained
astrometric results for six young stars. These results are listed in Table~\ref{tab:ORAM}. 
In order to compare the results of these studies with ours, we compute the weighted 
mean of absolute differences of proper motions
($\overline{|\Delta\mu_\alpha^{*}|}, \overline{|\Delta\mu_\delta|}$). 
The differences between the results 
of \citet{menten2007} and ours are ($0.3\pm0.1$, $0.7\pm0.1$)~mas~yr$^{-1}$, 
and a similar comparison for \citet{kounkel2017} yields
($0.2\pm0.1,0.2\pm0.1$)~mas~yr$^{-1}$. For the last
value, we have omitted the proper motions derived for [FRM2016]~184,
since the position changes are not consistent with a linear motion 
(see subsection \ref{invsou}).

 The results from the comparisons indicate that there are no significant 
differences with the measurements by \citet{kounkel2017}. On the other hand
the differences with \citet{menten2007}, are significant at levels of 3.0
and 7.0 times the errors. [FRM2016]~254 and [FRM2016]~378 are known to be part 
of binary systems \citep{petr1998,petr2008,duchene2018}, { and} orbital motions
will contribute to their final motions. The time separation between the
observations reported by \citet{menten2007} and ours is more than a decade, 
thus the difference between the proper motions of these sources, assumed lineal in both works,
can be due to the effects of orbital motions. To our knowledge,
neither [FRM2016]~400 or [FRM2016]~66 have been reported to be binary stars.
However, the discrepancies of measured proper motions for [FRM2016]~66
may suggest that it could also be part of a binary or multiple system.

\subsection{Trapezium region kinematics}\label{ssec:kin}

Proper motions of radio emitting YSOs in the core of the ONC and the BN/KL region have been 
used to study the global motions and internal kinematics of these regions
\citep{gomez2005,dzib2017}. These results used VLA observations spanning $\sim30$~years 
and were mainly limited by the angular resolution of the VLA observations. With the 
higher angular resolution afforded by the VLBA, these results
can be tested and significantly improved.

Using the VLBA proper motion values from Table~\ref{tab:RAM}, excluding sources 
[FRM2016]~18, 137, and 198 whose proper motions are significantly higher than the rest of the stars,
we estimate the mean value of the proper motion ($\overline{\mu}$).
In order to estimate the true proper motion dispersion ($\sigma_\mu$), we 
correct the observed dispersion for measurement uncertainty as suggested 
by \citet{jones1988}, i.e. 
$\sigma_\mu=\sigma_{\mu,Obs.}-\frac{1}{n}\sum\limits_{i=1}^n \mu_{\rm err, i}^2$.
For comparison, we have also estimated these
parameters for stars in the {\it Gaia} DR2 catalog. Restricting our
analysis to stars  located in the coverage of our radio observations,
with a parallax in the range from 2.4 to 2.6~mas (i.e., distances 
between 380 to 420 pc), and with Renormalised Unit Weight Error
(RUWE\footnote{The RUWE parameter is a quality indicator 
of \gaia\ fits. A RUWE value of 1.0  is expected
for sources whose motion is consistent with the motion of a  
single star. Larger values may suggest that the
source does not have good fit solutions, i.e., the
star may not be single. More information is given on the {\it Gaia} webpage: 
\url{ https://gea.esac.esa.int/archive/documentation/GDR2/Gaia_archive/chap_datamodel/sec_dm_main_tables/ssec_dm_ruwe.html}}) values smaller than 1.4, we used 12 stars for 
the calculations (see Figure~\ref{fig:VvsGPM}).  Finally,
as the astrometric precision from VLBA observations and 
the {\it Gaia} mission are comparable we estimated $\overline{\mu}$
and $\sigma_\mu$ combining these data sets. 
The results are shown in Table~\ref{tab:TK}, together with those obtained by 
\citet{dzib2017}. 

The averaged proper motions in the right ascension direction are consistent 
with a value of $\sim1.0$~mas~yr$^{-1}$ in the different data sets. 
For the declination direction both the VLBA and {\it Gaia}
DR2 data sets show motions near 0.0~mas~yr$^{-1}$, contrary 
to the larger value of $-0.8$~mas~yr$^{-1}$ obtained with the VLA.

The proper motion dispersions for the VLA and {\it Gaia} data sets are
close to a value of 1.0~mas~yr$^{-1}$ in both directions. Similar values 
were also found by \citet{jones1988} for optical stars covering a 
larger area of the ONC.  Recent values of proper motion dispersions
in the ONC have been estimated at optical and near infrared, to be 
($\sigma_\alpha,\sigma_\delta$)=($0.73\pm0.05,1.12\pm0.10$)~mas~yr$^{-1}$ using \gaia\ data
\citep[][]{kuhn2019}; 
($\sigma_\alpha,\sigma_\delta$)=($0.83\pm0.02,1.12\pm0.03$)~mas~yr$^{-1}$
using \textit{HST} and Keck II NIRC2 data \citep[][]{kim2019}; 
($\sigma_\alpha,\sigma_\delta$)=($0.89\pm0.03,1.21\pm0.04$)~mas~yr$^{-1}$
using \textit{HST} data \citep[][]{platais2020}. 
In the case of the VLBA data set the proper motion dispersion values are 
slightly larger. The large dispersions of VLBA proper motions
could reflect the multiplicity of systems related to detected radio sources. 
Because of its high angular resolution imaging capability, the observations with 
the VLBA can trace individual motions in these systems, while the observations 
at optical, infrared, and the low resolution observations at radio frequencies
will trace some intensity-weighted motion of the system.

Finally, we have also searched for signatures of expansion and contraction 
in the cluster following the analysis techniques of  
\citet{dzib2017} \citep[see also][]{rivera2015}. These authors used the mean 
values of the dot and cross products of a unit vector from the cluster
center toward star positions ($\mathbf{\hat{r}}={\mathbf{r}}/{||\mathbf{r}||}$) 
and the velocity vector ($\mathbf{v_*}$) of the individual stars. Large values 
in these products may indicate organized motion \citep[see][for a discussion]{rivera2015}. { For our analysis, the cluster center is defined as the average position of the stars with proper motion observed with the VLBA and the stars from {\it Gaia} with RUWE$\leq1.4$; to be ${(\alpha_0,\delta_0)=(5^{\rm h}35^{\rm m}16\rlap{.}^{\rm s}3,-5^\circ22'58'')}$\rlap{.}\footnote{The independent cluster center values for the used stars from the {\it Gaia} and VLBA catalog are 9'' away from this position.} }
From Table~\ref{tab:TK} we noticed that in all the cases
$\overline{\mathbf{v\cdot \hat{r}}}$ and $\overline{\mathbf{v}\times \mathbf{\hat{r}}}$ 
are consistent with 0.0~km~s$^{-1}$
within two times their uncertainties.  Even with our higher-resolution astrometry 
data, we find no strong evidence of rotation, contraction or expansion of the cluster,
in line with previous results.

\subsection{Individual sources}
\label{invsou}

While most of the stars show proper motions within $2\sigma$ of the mean motion
components,  four sources have motions well outside this range: 
 [FRM2016]~18, 137, 198 and 211.

Formally sources [FRM2016]~18, 137 and 198 exhibit the largest proper motions in our sample. Their proper motions are equivalent to velocities 
of 30, 80 and 25~km~s$^{-1}$).
Previous radio proper motions of source [FRM2016]~198 
were estimated by \citet{dzib2017},\footnote{Source named as 
VLA~J053514.66$-$052211.2 by these authors.} to be 
($-2.1\pm3.0$, $-0.9\pm1.4$)~mas~yr$^{-1}$, 
which are considerably lower than our values. Even with the lower 
resolution of the VLA \citet{dzib2017} would have detected such large 
proper motions. Since 
the VLBA proper motions are based on detections in only two epochs,
and allowing for the highly variable nature of the emission,
it is likely that [FRM2016]~198 is a binary system
with a separation of $\sim10$~mas, and the detections in the two epochs
correspond to different stars.  The large proper motion measured for
sources [FRM2016]~18 and 137 are intriguing, since they are well above
the proper motions exhibited for most of the stars in the ONC. However, 
with only two detections there is still the open possibility that they
correspond to different stellar components in multiple systems.
Future multi-epoch and deep VLBA observations, as those presented in this
work, will help to clarify the nature of the measured motions from the present observations.

[FRM2016]~211 was detected as a single source in the four observations.
In order to obtain a reduced $\chi^2=1.0$ in the proper motion fitting, 
we needed to add in quadrature values 1.2 and 2.4~mas to the right ascension
and declination position errors. These values are much larger than the 
expected systematic errors for VLBA observations, and, indeed, no other ONC star
shows such large uncertainties.  We suggest two explanations:
(1) the motion of [FRM2016]~211 is non-linear, or (2) we are 
detecting different stars among the observations.  Both explanations point 
to a binary nature of this system.

In our images we identify two compact radio sources in the 
direction of sources [FRM2016]~414 and [FRM2016]~177, which
have not been reported as binaries before.
Both sources show X-ray emission (see Paper I), and the spectral type
of [FRM2016]~414 is F8--K4 \citep{hillenbrand2013}.
The separation between the radio emission from stars in these system are 
0\rlap{.}$''$22 and 0\rlap{.}$''$023 for [FRM2016]~414 and [FRM2016]~177, respectively. 
With the current data it is not possible to constrain if they form
a gravitationally bound binaries, but further astrometric studies of 
these objects can determine this.  At high resolution, such a study will only 
be possible with VLBI techniques, as they are not detected at optical or NIR wavelengths.

In the direction of the VLA source [FRM2016]~2, we have 
single detections in epochs 3 and 4; i.e. with a time baseline one year. 
The separation between the radio sources is 20.6~mas. Assuming that 
both radio sources are the same star would imply a large proper motion of 
20.6~mas~yr$^{-1}$ ($\simeq\,$39~km~s$^{-1}$). However, the Gaia source 
associated with this source has a proper motion of 
($\mu_\alpha*,\mu_\delta$)=($1.64\pm0.08,0.07\pm0.07$)~mas~yr$^{-1}$, 
inconsistent with a source with fast motion. Furthermore, extrapolating 
the 2015.5 position of the {\it Gaia} source to epochs 3 (2017.8) and 4 (2018.8), 
find it to be separated $22.6\pm0.3$~mas and $1.6\pm0.3$~mas from the radio sources detected
in epochs 3 and 4, respectively, suggesting that only the source detected in 
epoch 4 is probably directly related to the optical source. Our conclusion for [FRM2016]~2 
is that the two radio sources detected in the two different epochs are not the same.

Toward the VLA source [FRM2016]~184, we detected single sources in all four epochs.
The proper motion using the four epochs is
($\mu_\alpha*,\mu_\delta$)=($0.71\pm0.28,-1.71\pm0.91$)~mas~yr$^{-1}$,
where we have added { in quadrature} additional values of 0.33 and 1.66~mas in right 
ascension and declination, respectively, to the position errors to obtain
a $\chi^2=1$ in our fit. The detections in epochs 2 and 3 are
consistent within errors as expected since the separation between the epochs is only
one day, suggesting that the source in these epochs is the same. 
The measured proper motion between epochs 1 and 2+3 is
($\mu_\alpha*,\mu_\delta$)=($0.34\pm0.19,-2.95\pm0.26$)~mas~yr$^{-1}$,
while the proper motion from epochs 2+3 and 4 is 
($\mu_\alpha*,\mu_\delta$)=($1.46\pm0.20,1.11\pm0.18$)~mas~yr$^{-1}$.
Furthermore, these motions are different from the proper motions reported
by \citet{kounkel2017}, who also used VLBA observations. The motion related
to [FRM2016]~184 deserves further investigation to clarify its unusual motion.

$\theta^{1}$~Ori~A is a hierarchical triple system known to have a total mass of 
$\sim20$~M$_\odot$.  At optical wavelengths, the brightest component is $\theta^{1}$~Ori~A$_1$,
which is a tight binary itself \citep[P=65.433 days,][]{lloyd1999,bondar2000}, 
composed of a  $\approx 15$~M$_\odot$, massive star,
\citep{weigelt1999, schertl2003, simondiaz2006,nieva2014} and a T-Tauri star of 
$\approx 2.5$~M$_\odot$ \citep[][]{bossi1989}.  At an angular distance of 0\rlap{.}$''$18 from 
the tight binary, there is the 4~M$_\odot$ star $\theta^{1}$~Ori~A$_2$ \citep{petr1998}. 
NIR interferometry data taken from 1994 to 2010 and analyzed by \cite{Grellmann2013}  
indicated linear movement of A$_2$ relative to A$_1$ suggesting unbound motion. 
However, adding new NIR interferometry data taken between 2016 and 2018, the relative motions determined by
\citet{gravity2018} strongly suggest that $\theta^{1}$~Ori~A$_2$ is gravitationally bound to the 
tight binary $\theta^{1}$~Ori~A$_1$. 
As discussed before, the optical source
seen by \gaia\ is related to component A$_1$, and the 
radio source is related to component A$_2$.
The total differences between the VLBA and \gaia\ 
proper motions is $4.3\pm0.2$~mas~yr$^{-1}$, which at the distance of 
the ONC  is equivalent to a velocity of $8.2\pm0.4$~km~s$^{-1}$. 
Assuming, in a simplifying estimate, a circular orbit for the system,
the escape velocity of A$_2$ from A$_1$, with a total mass of 
18~M$_\odot$, is 21~km~s$^{-1}$. This would indicate indeed that 
the system is bound, as the relative velocity
between the components would be significantly smaller than the escape velocity.

\section{Conclusions}

We have imaged 126 compact nonthermal radio sources near  
the Trapezium in the ONC using data acquired with VLBA observations over a three year period.
The positions of 34 radio detected sources were found within $0\rlap{.}''2$
of a \gaia star.  
%The positions of all detected 
%sources were measured and \gaia counterparts were found within $0\rlap{.}''2$
%of {34} radio sources.  
Most of the \gaia sources (23) are 
well separated ($>4$~mas) from the associated radio source, indicating that the 
optical and the radio source are not the same star.  We argue that
the radio sources could be lower-mass companions of the stars seen
by {\it Gaia}.  For the remaining 11 cases, the separation 
is $<1.6$~mas and could indicate that both telescopes observe the
same stellar sources. 
We find mean separations of ${-0.32\pm0.15\pm0.16}$~mas and ${0.99\pm0.10\pm0.16}$~mas
in right ascension and declination, respectively.  
The stellar position difference in declination between both catalogs is significant
at a level of ${5\sigma}$. Its origin is still uncertain, but it could conceivably be 
due to unidentified binaries in the sample. 
 For three targets, { two separate components were identified in the images}, further expanding the discovery space for previously unknown multiple systems in the ONC.

$\overline{\Delta\alpha}={-0.32\pm0.15 ({\rm stat.}) \pm 0.16 ({\rm sys.})}$~mas and 
$\overline{\Delta\delta}={0.99\pm0.10 ({\rm stat.}) \pm 0.16 ({\rm sys.})}$~mas,

Radio proper motions were estimated for  23 YSOs with
accuracies of $\approx0.1$~mas~yr$^{-1}$, similar to {\it Gaia} accuracy. 
Within 1.6~mas, five of them have a counterpart in the \gaia\ catalog,
and by comparing their proper motions from both catalogs, we found differences 
of $-0.025\pm0.080$ and $-0.070\pm0.090$~mas~yr$^{-1}$ in right ascension 
and declination, respectively. 

By combining proper motions  from \gaia\ and VLBA
we have improved the values of the global motions and the 
kinematic of the ONC core.  The global proper motion
and velocity dispersion are
($\mu_\alpha^{*},\mu_\delta$)=($1.20\pm0.09,0.18\pm0.09$)~mas~yr$^{-1}$ and
($\sigma_{\mu_{\alpha}^{*}},\sigma_{\mu_\delta}$)=($0.84\pm0.09,1.30\pm0.09$)~mas~yr$^{-1}$, 
respectively. The search of ordered motion through vector products 
$\overline{\mathbf{v\cdot \hat{r}}}$ and $\overline{\mathbf{v}\times \mathbf{\hat{r}}}$ 
show that the obtained values are consistent with a value of 0.0~km~s$^{-1}$ 
within two times the errors. These results do not show indications 
of expansion/contraction or rotation of the young stellar cluster.

\acknowledgments 
{\small
The National Radio Astronomy
Observatory is a facility of the National Science Foundation
operated under cooperative agreement by Associated 
Universities, Inc.

This work has made use of data from the European Space 
Agency (ESA) mission {\it Gaia}
(\url{https://www.cosmos.esa.int/gaia}), processed by the 
{\it Gaia} Data Processing and Analysis Consortium (DPAC,
\url{https://www.cosmos.esa.int/web/gaia/dpac/consortium}). 
Funding for the DPAC has been provided by national 
institutions, in particular the institutions
participating in the {\it Gaia} Multilateral Agreement.
}
%\end{acknowledgements}
%\facility{facility ID}

\facilities{VLBA} 
%\software{}

%\newpage
\appendix

\section{VLBA and \gaia\ positions differences of extragalactic sources around the ONC \label{apend:A}}

We searched for extragalactic sources in a radius of $5^{\circ}$ around the ONC that 
have compact radio emission and that are also in the \gaia\ catalog. We found eight 
sources meeting these requirements and list them in Table~\ref{tab:posQ}. Their most 
recently determined radio positions and their uncertainties were taken
from the AstroGeo
catalog {\textit{rfc\_2020c}}\footnote{\url{http://astrogeo.org/rfc/}} and are shown in Table~\ref{tab:posQ}. 
The difference between the position of each radio source and that of its 
\gaia\ counterpart is also shown in Table ~\ref{tab:posQ}.

Previous comparisons of the positions compact extragalactic objects based on data from 
Very Long Baseline Interferometry and the Gaia satellite of have found that around 10\% 
of them have significant offsets between them \citep{petrov2017,petrov2019}. 
These differences have a physical origin and are mainly due the spatially different appearance 
of optical and radio jets, which both evolve with time \citep[see discussion by][]{petrov2019}. 
Statistically, we expect that 0.8 sources from our sample are 
part of this group. In our sample, source J0552$-$0727
has a significant separation of 28.4~mas, and was thus omitted from our analysis.

The statistical analysis of the separation of the remaining sources yield mean separations
of ($\Delta\alpha$,$\Delta\delta$)=(${-0.6\pm0.7}$, ${0.0\pm0.6}$)~mas that have
standard deviations of ($\sigma_{\Delta\alpha}$,$\sigma_{\Delta\delta}$)=(1.8,1.4)~mas. 
The mean separation of our sample is smaller than the median separation found by \citet{petrov2017} 
for VLBI and Gaia DR1 positions. However, the large dispersion { in the} values may suggest %that there are 
differences between the radio and \gaia\ reference frames.

\begin{deluxetable*}{crrrrrcccc}
\renewcommand{\arraystretch}{0.9}
%\tablenum{1}
\tablecaption{Positions at radio frequencies of extragalactic sources around the ONC and their separations from \gaia. 
\label{tab:posQ}}
\tablewidth{0pt}
\tablehead{
\colhead{} & 
\colhead{} & 
\colhead{$\sigma_\alpha$} & %_{\rm err}$} 
%\colhead{$\alpha_{\rm err}$} & 
\colhead{} & 
\colhead{$\sigma_\delta$} & %_{\rm err}$} 
\colhead{}&\colhead{$\Delta\alpha$}&\colhead{$\Delta\delta$}\\ 
\colhead{Name}&  \colhead{$\alpha_{{\rm J}2000}$} & \colhead{$\mu$s} &\colhead{$\delta_{{\rm J}2000}$}  & \colhead{$\mu$as} &\colhead{{\it Gaia} ID (DR2)}&\colhead{(mas)}&\colhead{(mas)}
}
%\decimalcolnumbers
\startdata
\hline
J0539$-$0514 &\rahms{05}{39}{59}{937139} & 10 & \decdms{-05}{14}{41}{30061} &  280 & 3023327569572290688 & $-0.4\pm0.2$ & $-0.2\pm0.3$ \\
J0529$-$0519 &\rahms{05}{29}{53}{533500} & 7  & \decdms{-05}{19}{41}{61733} &  160 & 3209472860931863424 & $-2.0\pm1.3$ & $-0.9\pm1.5$ \\
J0541$-$0541 &\rahms{05}{41}{38}{083368} & 7  & \decdms{-05}{41}{49}{42846} &  110 & 3017106773301050240 & $+0.8\pm1.5$ & $+0.9\pm1.4$ \\
J0532$-$0307 &\rahms{05}{32}{07}{519331} & 8  & \decdms{-03}{07}{07}{03649} &  190 & 3216726171637867136 & $+0.7\pm1.0$ & $+2.9\pm1.0$ \\
J0545$-$0539 &\rahms{05}{45}{23}{358039} & 17 & \decdms{-05}{39}{37}{83964} &  350 & 3022323749816621568 & $-4.2\pm0.5$ & $-0.7\pm0.8$ \\
J0522$-$0725 &\rahms{05}{22}{23}{196758} & 52 & \decdms{-07}{25}{13}{48025} & 1180 & 3207290781323060096 & $+1.0\pm0.8$ & $-1.5\pm1.2$ \\
J0517$-$0520 &\rahms{05}{17}{28}{110159} & 8  & \decdms{-05}{20}{40}{84120} &  190 & 3208721928848872576 & $-0.1\pm0.3$ & $+0.2\pm0.3$ \\
J0552$-$0727 &\rahms{05}{52}{11}{376231} & 9  & \decdms{-07}{27}{22}{51824} &  250 & 3018834797558565376 & $+2.4\pm1.0$ & $-28.3\pm1.1$ \\
\enddata
\tablecomments{Columns are (left to right): Source name, J2000 positions in right ascension and declination, both with uncertainties, {\it Gaia} ID, and separations between the Gaia and VLBA positions in both coordinate directions.
{ The separations are defined as: $\Delta\alpha=(\alpha_{\rm VLBA}-\alpha_{\rm \gaia})\cdot\cos{\delta}$ and $\Delta\delta=(\delta_{\rm VLBA}-\delta_{\rm \gaia})$.}
}
\end{deluxetable*}

\section{Stars with good astrometry in \gaia\  and inside the area of our radio study \label{apend:B}}

 In this appendix we give the list of the stars in the \gaia\ catalog used for the estimation of values in Table~\ref{tab:TK}. The criteria used to compile this list were: to be in the same area as the primary beam of the VLBA observations, the measured parallax suggesting a distance between 380 and 420 pc, and to have a RUWE parameter $\leq1.4$.
These stars are listed in Table~\ref{tab:posS} with their positions and proper motions.

\begin{deluxetable*}{ccccccccccc}
\renewcommand{\arraystretch}{0.9}
%\tablenum{1}
\tablecaption{Stars in the \gaia\ catalog used for the estimation of values in Table~\ref{tab:TK}.
\label{tab:posS}}
\tablewidth{0pt}
\tablehead{
\colhead{} & 
\colhead{Sp.} & 
\colhead{} &
\colhead{}&
\colhead{} & 
\colhead{$\sigma_\alpha$} & %_{\rm err}$} 
%\colhead{$\alpha_{\rm err}$} & 
\colhead{} & 
\colhead{$\sigma_\delta$} & 
%_{\rm err}$} 
\colhead{$\mu_\alpha*$}&\colhead{$\mu_\delta$}\\ 
\colhead{Name}&
\colhead{Type}&
\colhead{{\it Gaia} ID (DR2)}&
\colhead{RUWE} &
\colhead{$\alpha_{{\rm J}2015.5}$} & \colhead{$\mu$s} &\colhead{$\delta_{{\rm J}2015.5}$}  & \colhead{$\mu$as} &\colhead{(mas yr$^{-1}$)}&\colhead{(mas yr$^{-1}$)}
}
%\decimalcolnumbers
\startdata
\hline
$\theta^1$ Ori F&B8&3017364063331140224&1.0&\rahms{5}{35}{16}{732576}&4& \decdms{-5}{23}{25}{22448}&60&$5.45\pm0.11$&$-1.77\pm0.10$\\
$\theta^1$ Ori B&B1V&3017364132049943680&1.0&\rahms{5}{35}{16}{135316}&3& \decdms{-5}{23}{06}{76466}&49&$1.46\pm0.11$&$1.03\pm0.10$\\
V2325 Ori&M0&3017364063330467072&1.4&\rahms{5}{35}{18}{208086}&4& \decdms{-5}{23}{35}{90580}&52&$1.32\pm0.13$&$-0.00\pm0.10$\\
$\theta^1$ Ori E&G2IV&3017364127743288704&1.2&\rahms{5}{35}{15}{773584}&4& \decdms{-5}{23}{09}{87097}&72&$1.61\pm0.12$&$1.23\pm0.11$\\
V348 Ori&G8-K1&3017364127743288320&1.2&\rahms{5}{35}{15}{636427}&3& \decdms{-5}{22}{56}{43502}&43&$1.08\pm0.12$&$1.45\pm0.10$\\
Brun 633&A4-A7&3017365880089961728&1.0&\rahms{5}{35}{19}{139597}&2& \decdms{-5}{20}{38}{72779}&35&$0.14\pm0.08$&$1.20\pm0.07$\\
GMR G &K2&3017364127743299328&1.2&\rahms{5}{35}{17}{952386}&3& \decdms{-5}{22}{45}{4353}&38&$3.77\pm0.10$&$2.34\pm0.08$\\
GMR F&K0&3017364162103039104&1.2&\rahms{5}{35}{18}{372793}&3& \decdms{-5}{22}{37}{42811}&38&$2.13\pm0.10$&$0.78\pm0.08$\\
Brun 676&K3&3017365880089976064&1.0&\rahms{5}{35}{22}{265125}&2& \decdms{-5}{20}{29}{26229}&34&$1.56\pm0.07$&$-0.36\pm0.07$\\
MR Ori&A2:Vv&3017364372568073472&0.9&\rahms{5}{35}{16}{979192}&2& \decdms{-5}{21}{45}{31264}&33&$0.90\pm0.07$&$1.56\pm0.06$\\
LV Ori&K1V&3017364028971010432&1.2&\rahms{5}{35}{12}{601447}&2& \decdms{-5}{23}{44}{13115}&38&$2.57\pm0.08$&$-1.36\pm0.08$\\
V1326 Ori&K8&3017363994611276032&1.1&\rahms{5}{35}{09}{769407}&3& \decdms{-5}{23}{26}{89052}&37&$1.02\pm0.12$&$0.56\pm0.09$\\
%Parenago 1754&M1&3017363612347069440&1.3&\rahms{5}{35}{6}{519363}&2& \decdms{-5}{24}{41}{39704}&32&$3.56\pm0.10$&$1.33\pm0.07$
\enddata
\tablecomments{Columns are (left to right): Source name, Spectral type, {\it Gaia} ID in the DR2 catalog, Gaia DR2 positions in right ascension and declination given in the epoch J2015.5, both with uncertainties, and proper motions in both coordinate directions.
}
\end{deluxetable*}

\newpage
\bibliographystyle{aa}
\bibliography{references}

\end{document}